\documentclass[preprint,pre,showpacs,preprintnumbers,amsmath,amssymb]{revtex4}

\usepackage{dcolumn}
\usepackage{bm}
\usepackage{graphicx,subfigure,endnotes}
\usepackage{float}

\newcommand{\deldel}{\nabla \times \nabla \times}
\newcommand{\Ev}{{\bf E}}
\newcommand{\Er}{{\bf E}({\bf r})}
\newcommand{\Enr}{{\bf E}_0({\bf r})}
\newcommand{\Epr}{{\bf E}({\bf r}^\prime)}
\newcommand{\Gv}{{\bf G}}
\newcommand{\bfm}[1]{\mbox{\boldmath ${#1}$}}
\newcommand{\epeff}{{\bfm\varepsilon}_e}
\newcommand{\Hv}{{\bf H}}
\newcommand{\Iv}{{\bf I}}
\newcommand{\Lv}{{\bf L}}
\newcommand{\Fv}{{\bf F}}
\newcommand{\Pv}{{\bf P}}

\newcommand{\rv}{{\bf r}}
\newcommand{\Rv}{{\bf R}}
\newcommand{\qv}{{\bf q}}
\newcommand{\tv}{{\bf t}}
\newcommand{\Tr}{\mbox{Tr}}
\newcommand{\Imag}{\mbox{Im}}
\newcommand{\Ind}{{\cal I}}

\newcommand{\rhatrhat}{\hat{\rv}\hat{\rv}}

\begin{document}

\title{Effective Dielectric Tensor for Electromagnetic Wave
Propagation in Random Media}

\author{Mikael C. Rechtsman$^1$} \author{Salvatore Torquato$^{1,2,3,4,5}$}

\email{torquato@electron.princeton.edu}

\affiliation{$^1$Department of Physics, Princeton University,
Princeton, New Jersey, 08544} \affiliation{$^2$Department of
Chemistry, Princeton University, Princeton, New Jersey, 08544}
\affiliation{$^3$Program in Applied and Computational Mathematics, Princeton, New Jersey, 08544}
\affiliation{$^4$Princeton Institute for the Sceince and Technology
of Materials, Princeton, New Jersey, 08544} \affiliation{$^5$Princeton Center
for Theoretical Physics, Princeton, New Jersey, 08544} 
\date{\today}

\begin{abstract}
We derive exact strong-contrast expansions for the effective
dielectric tensor $\epeff$ of electromagnetic waves propagating in a
two-phase composite random medium with isotropic components explicitly in terms
of certain integrals over the $n$-point correlation functions of the medium.  Our focus is the
long-wavelength regime, i.e., when the wavelength is much larger than
the scale of inhomogeneities in the medium. Lower-order truncations of
these expansions lead to approximations for the effective dielectric
constant that depend upon whether the medium is below or above
the percolation threshold.  In particular, we apply two- and
three-point approximations for $\epeff$ to a variety of different
three-dimensional model microstructures, including dispersions of hard
spheres, hard oriented spheroids and fully penetrable spheres as well as  Debye random media,
the random checkerboard, and power-law-correlated materials.  We
demonstrate the importance of employing $n$-point correlation
functions of order higher than two for high dielectric-phase-contrast ratio.
We show that disorder in the microstructure results in an imaginary
component of the effective dielectric tensor that is directly related
to the {\it coarseness} of the composite, i.e., local
volume-fraction fluctuations for infinitely large windows.  The
source of this imaginary component is the attenuation of the coherent
homogenized wave due to scattering. We also remark on whether
there is such attenuation in the case of a two-phase medium
with a quasiperiodic structure.
\end{abstract}

\maketitle

\section{Introduction}

The problem of determining the effective dielectric tensor and other
mathematically equivalent properties of random  media dates
back to the classic work of Maxwell \cite{Maxwell}.  Calculation of
the effective dielectric tensor of disordered composite materials is
essential for a wide range of applications, including remote sensing
(e.g., of terrain, vegetation, etc.) \cite{RemoteSensing}, the study of
wave propagation through turbulent atmospheres \cite{Turbulent}, 
active manipulation of composites \cite{Me02}, and as a probe of artificial
materials \cite{Zhuck}. The effective dielectric tensor at long
wavelengths plays a particularly important role in the study of
electrostatic resonances \cite{Mc80,Brosseau}.

This paper is concerned with the calculation of the effective
dielectric tensor $\epeff$ of a two-phase dielectric random medium associated
with electromagnetic wave propagation in the long wavelength regime,
i.e., when the wavelength is much larger than the scale of
inhomogeneities in the medium. The complementary regime, 
in which the wavelength is much smaller than the inhomogeneities, may be studied
numerically using a ray-tracing technique based on geometrical optics \cite{Zohdi}.
 The purely static (as opposed to the
full dynamic) problem, on which Maxwell's work was centered, can be
considered to be a special case of the present work in the limit of
infinite wavelength (or zero frequency).  Although there has been
extensive previous work on the dynamic problem
(see Refs. \onlinecite{Mc99} and \onlinecite{Sheng} and references therein), the vast majority 
of studies that attempt to relate $\epeff$ to the microstructure
employ only two-point correlation information \cite{TK,Zhuck,Klusch} 
(with a few exceptions, see, e.g.,  Ref. \onlinecite{Mackay}).  
On the other hand, there has been a great deal of work
on the static problem incorporating three-point and higher-order
correlation functions; see, for example Refs. \onlinecite{Beran,
Milton,Mc81,Torquato1,Torquato3point1,RHM}, and references therein.

Here we present, for the first time, explicit closed-form series
expansions for the effective dielectric tensor of two-phase random
media in three dimensions in terms of certain integrals of
$n$-point correlation functions for the dynamic problem in the long-wavelength regime.  The
approach used follows directly from one given originally for the
purely static problem in any dimension in Ref. \onlinecite{Torquato1}, and expanded
upon in Ref. \onlinecite{RHM}.  An advantageous feature of this formalism
is that it gives rise to expansion parameters involving
the dielectric constants that yield very good convergence
properties even for high phase-contrast ratio.  The technique is
called the {\it strong-contrast expansion} for this reason.  Tsang and
Kong \cite{TK} have employed a similar method, but only up to the
two-point level and  they provide no justification
for or limitations on the class of microstructures and phase-contrast
ratios for which their two-point approximation  is applicable.  We restrict ourselves to
considering component materials that are themselves isotropic but
that generally possess complex-valued dielectric constants.
As we will demonstrate, lower-order truncated forms of these
expansions can serve as useful approximations of the effective
dielectric tensor.  Although the formalism is applied here to
electromagnetism in three dimensions, it is straightforwardly
generalizable to a class of other vector fields in arbitrary
dimension \cite{RHM}.

 We obtain new approximations for $\epeff$ by truncating the
exact strong-contrast expansions at the two-point and three-point
levels that depend upon whether the medium is below or above its
percolation threshold.  Using these two-point approximations, we
estimate $\epeff$ of dispersions  of hard spheres, hard oriented spheroids and fully
penetrable spheres as well as  Debye random media, random checkerboard, and
power-law-correlated materials.  We use a three-point approximation to
evaluate the effective dielectric constant in the case of the
fully-penetrable-sphere model in order to establish the importance of
three-point information at high phase-contrast ratio and volume
fraction, and assess the accuracy of the two-point approximations.  
For many of the model microstructure considered here, the 
explicit forms of the corresponding $n$-point correlation functions follow from the general
representation formalism  of Torquato and Stell \cite{TorquatoStell}.

A significant qualitative difference between the purely static problem
and the dynamic problem is that in the latter, the effective
dielectric tensor may be complex even if the component materials have
purely real dielectric constants.  This results from incoherent
scattering of the incident wave.  The imaginary component of the
effective dielectric tensor has special significance to remote sensing
applications, as it plays a key role in, for example, the calculation
of backscattering coefficients \cite{TK}.  Although the expansions we
derive here are applicable in general to materials with complex
dielectric constants, we will apply them to cases in which the
material phases have real dielectric constants.  We do this because in
this regime, disordered media will in general yield nonnegative
imaginary contributions to the effective dielectric tensor purely due
to scattering, not absorption.  The physical manifestation of
this can be understood in the context of optical or ultrasonic
transmission experiments \cite{BallisticDiffusive,
BallisticDiffusive2}, in which a wave pulse propagates through a
finite-sized slab composed of a disordered material, and the
transmitted signal is measured.  The transmitted field is observed to
be a superposition of a coherent pulse representing the incident
signal and uncorrelated noise due to random scattering.  The
attenuation of the incident pulse is a consequence of this
scattering. Since periodic media propagate waves without any loss
(and thus with purely real effective dielectric tensors), they do not
exhibit imaginary parts in their effective dielectric tensors.  In the
case of statistically homogeneous and isotropic media in particular,
we show here that the leading-order contribution to the imaginary
component of the effective dielectric constant is directly related to
the {\it coarseness} $C_\infty$ of the composite for an infinitely
large window.  The coarseness is defined as the standard deviation of
the volume fraction of a given phase, in some observation ``window''
within the composite divided by the total volume fraction of that
phase \cite{RHM,LuTorquato}.  It has recently been suggested that
local density fluctuations provides a crude measure of disorder of
a system \cite{StillingerTorquato,Footnote}.  In the language of
Ref. \onlinecite{StillingerTorquato}, a {\it hyperuniform} system is one in which
$C_\infty=0$, which has been shown to always be the case for periodic
systems, and therefore  is consistent with the fact that
$\Imag\left[\epeff\right]=0$ for such media. Note that a non-zero imaginary component of the effective dielectric constant
can be directly associated with a mean free path of waves propagating through the given medium \cite{Sheng}.

In Section II, we introduce our formalism, derive the strong-contrast
expansion for the effective dielectric tensor, and discuss the
approach towards applying these techniques at the three-point level.
In Section III, we discuss a variety of model microstructures not previously
studied in the present context and
their associated two-point correlation functions.  In Section IV, we present results
for the model microstructures discussed in Section III, showing volume
fraction and dielectric-contrast ratio dependence at the two-point
level, and demonstrating the importance of three-point calculations at
sufficiently high phase contrasts and volume fractions.  In Section V, we discuss
conclusions based on our results.  Appendix A presents a
generalization of our formalism, up to the two-point level, wherein
the expansion is carried out with an arbitrary ``reference"
or ``comparison" material (as described in Section II.A).
discussed hence).  Appendix B gives a short proof of why periodic
media, at low frequencies, must give rise to effective dielectric
tensors with no imaginary component.  Finally, in Appendix C,
we simplify the key two-point integral for
statistically homogeneous but anisotropic media
with azimuthal symmetry, an example of which is
a dispersion of hard oriented spheroids.

\section{Theory}

Here we extend the formalism for determining strong-contrast
expansions of the purely static effective dielectric tensor  
in arbitrary space dimensions  \cite{RHM,Torquato1} to the dynamic
case in $\mathbb{R}^3$ when the wavelength is much larger than the inhomogeneity
length scale.  This method is
based on finding solutions of certain integral equations using the
method of Green's functions.

\subsection{Strong-Contrast Expansions}

Following Ref. \onlinecite{RHM} (which gives greater
detail than Ref. \cite{Torquato1}), we begin by considering a macroscopically
large ellipsoidal sample composite material in $\mathbb{R}^3$ composed
of two phases, labeled 1 and 2, which is itself embedded in a
homogeneous ``reference" (or ``comparison) material of dielectric constant $\varepsilon_0$.  We choose a
specific macroscopic shape to call attention to the fact that the
average fields in the problem are shape dependent.  An advantage of
this formalism is that it eliminates the shape dependence of the
effective dielectric tensor.  The length scale of inhomogeneities
within the composite is assumed to be much smaller than the shape
itself.  The two phases have dielectric constants $\varepsilon_1$ and
$\varepsilon_2$, and we define indicator functions in the following way:
\begin{equation}
\Ind^{(p)}(\rv) = \left\{ \begin{array}{l}
1, \mbox{ if } \rv \mbox{ lies in phase } p \\
0,  \mbox{ otherwise}\\ 
\end{array}\right .
\end{equation}
for $p=1,2$.  Thus, the volume fractions of the phases are
$\langle\Ind^{(p)}(\rv)\rangle=\phi_p$ and $\phi_q=1-\phi_p$ ($p\neq
q$).  We may therefore write $\varepsilon (\rv)=\varepsilon_1
\Ind^{(1)}(\rv)+\varepsilon_2 \Ind^{(2)}(\rv)$ in the composite.  Note
that the dielectric constants $\varepsilon_1$ and $\varepsilon_2$ may be
complex.

When solved for the electric field, time-harmonic Maxwell's equations reduce to
\begin{equation}
\deldel \Er - \varepsilon(\rv)\left(\frac{\omega}{c}\right)^2\Er = 0,
\end{equation}
with $\Ev$ the electric field, $\varepsilon(\rv)$ the reduced dielectric
constant, $\omega$ the frequency of the time-harmonic solution, and
$c$ the speed of light.  We assume here that $\mu/\mu_0=1$, and that
the component dielectric materials are themselves isotropic.
We can rewrite this homogeneous, linear equation in a form that is
suggestive of perturbation theory:
\begin{equation}
\deldel \Er - \varepsilon_q\left(\frac{\omega}{c}\right)^2\Er =
(\varepsilon(\rv)-\varepsilon_q)\left(\frac{\omega}{c}\right)^2\Er,\label{perturb1}
\end{equation}
where we have taken the comparison material to be one of the
phase materials, i.e., $\varepsilon_0=\varepsilon_q$, where $q$ is 
either 1 or 2.  We employ this choice for simplicity
here, but this is not essential; in fact, depending on the details of
the structure in question, it may improve convergence to choose 
a different comparison material \cite{RHM,Eyre1,Ph03}.  We discuss
this point in further detail in Appendix A.

As a shorthand, we will define
$k_q^2\equiv\sigma_q\equiv\left(\frac{\omega}{c}\right)^2\varepsilon_q$,
for any phase, $q=1 \mbox{ or } 2$.  In order to solve this equation
for an arbitrary structure via perturbation theory, we require a
Green's function for the operator on the left side of
Eq. (\ref{perturb1}), which must therefore be a tensor.  This is the
dyadic Green's function.  In three-dimensional spherical coordinates,
it is given by
\begin{equation}
\Gv(\rv,\rv^\prime) = -\frac{{\bf I}}{d\sigma_q}\delta(\rv-\rv^\prime)
+ G_1(\rv-\rv^\prime){\bf I} +
G_2(\rv-\rv^\prime)\hat{\rv}\hat{\rv},\label{GF}
\end{equation}
where
\begin{eqnarray*}
G_1(\rv-\rv^\prime) = (-1 + ik_q r + k_q^2 r^2) \frac{e^{ik_q r}}{4\pi
k_q^2 r^3},\\ G_2(\rv-\rv^\prime) = (3 - 3ik_q r - \sigma_q r^2)
\frac{e^{ik_q r}}{4\pi k_q^2 r^3}\\,
\end{eqnarray*}
with $r=|\rv-\rv^\prime|$, $\hat{\rv}$ is a unit vector directed from
$\rv^\prime$ towards $\rv$, and $\Iv$ is the unit tensor in three
dimensions.  In the limit $k_q\rightarrow 0$, we recover the static
result \cite{RHM} in three dimensions.  Note the delta function in
this expression; this is the dipole ``source'' of the radiation; its
coefficient is dependent on the shape of the ``exclusion volume''
around this source.  For the coefficient shown, the exclusion volume
must be spherical in shape.

The Green's function satisfies the following partial differential
equation:
\begin{equation}
\deldel \Gv (\rv, \rv^\prime)-\sigma_q \Gv (\rv, \rv^\prime) = {\bf
I}\delta(\rv-\rv^\prime).
\end{equation}
This implies that we can write the electric field as
\begin{equation}
\Er = \Enr + \int d\rv^\prime \Gv (\rv, \rv^\prime)
\left[\sigma(\rv)-\sigma_q\right]\Epr.
\end{equation}
We can express this integral equation more compactly in linear
operator form:
\begin{equation}
\Ev = \Ev_0 + \Gv\Pv,
\end{equation}
where we define the polarization vector field as
\begin{equation}
\Pv=\left[\sigma(\rv)-\sigma_q\right]\Ev.  
\end{equation}
The next step is to extract the delta function contribution from the
Green's function solution.  In doing so, we obtain a new field, $\Fv$,
the {\it cavity intensity field}, which is directly proportional to
$\Ev$.  The resulting integral equation is
\begin{equation}
\Fv = \Ev_0 + \Hv\Pv,\label{selfconsistent}
\end{equation}
where $\Hv$ is the principle value of the Green's function in
Eq. (\ref{GF}), namely
\begin{equation}
\Hv(\rv-\rv^\prime)=G_1(\rv-\rv^\prime){\bf I} +
G_2(\rv-\rv^\prime)\hat{\rv}\hat{\rv}.\label{H-definition}
\end{equation}

Here, we have defined
\begin{equation} 
\Fv({\bf r}) = \left[{\bf I} + \frac{{\bf
\sigma}(\rv)-\sigma_q}{d\sigma_q}\right]\cdot \Er,
\end{equation}
where ${\bf I}$ is the unit dyadic tensor.  Now, $\Pv$ and $\Fv$ are
directly related:
\begin{equation}
\Pv(\rv) = \frac{\sigma (\rv)-\sigma_q}{{\bf I} + \displaystyle
\frac{{\bf I}}{d\sigma_q}\left[{\bf
\sigma}(\rv)-\sigma_q\right]}\Fv(\rv).
\end{equation}
or, implicitly defining $\Lv$,
\begin{equation}
\Pv(\rv) = \Lv(\rv)\cdot\Fv(\rv).
\end{equation}
Note that $\Lv$ is an isotropic tensor.  Instead of using the standard
definition of the effective dielectric tensor, $\langle {\bf D}\rangle
=\varepsilon_{e} \cdot \langle \Ev\rangle $, we may equally well use the above
equation.  Thus, implicitly defining $\Lv_{e}$, we have
\begin{equation}
\langle \Pv(\rv)\rangle = \Lv_{e}\cdot\langle \Fv(\rv)\rangle,
\label{homogenization}
\end{equation}
where $\langle .\rangle $ denotes the ensemble (volume) average
(assuming ergodicity).  We may write ${\bf L}_{e}$ explicitly at this
point.  It is given by
\begin{equation}
\Lv_{e}=L_{e}\Iv=\frac{{\bf \sigma}_{e} -\sigma_q}{{\bf I} +
\displaystyle\frac{{\bf I}}{d\sigma_q}\left[{\bf
\sigma}_{e}-\sigma_q\right]},
\end{equation}
where ${\bf \sigma}_{e} = \left(\frac{\omega}{c}\right)^2{\bf
\varepsilon}_{e}$, with ${\bf \varepsilon}_{e}$ the effective dielectric
tensor of the composite.

We now have an equation that defines the effective dielectric tensor.
The last step involves eliminating the background field, $\Ev_0$.
This is done because, as is known from electrostatics, the
relationship between the applied field and the average fields ($\Pv$
and $\Lv$, for example) is shape-dependent.  Thus, eliminating $\Ev_0$
results in an effective dielectric constant that is independent of the
shape of the macroscopic ellipsoid.  It has been demonstrated that the
elimination $\Ev_0$ thus leads to integrals of correlation functions
within the formalism that are absolutely convergent \cite{RHM,Torquato1}.

We define here a new tensor operator ${\bf S}$ in the following way:
\begin{equation}
{\bf S} = \Lv\left[\Iv-\Lv\Hv\right]^{-1}.
\end{equation}
It follows directly that $\langle \Pv\rangle =\langle {\bf S}\rangle
\Ev_0$, and we may thus eliminate $\Ev_0$ from
Eq. (\ref{selfconsistent}).  Upon taking the ensemble average of the
latter, we may write $\langle \Fv\rangle $ in terms of $\langle
\Pv\rangle$, the first few terms of which are given by:
\begin{equation}
\langle\Fv(\rv_1)\rangle=\frac{\langle\Pv (\rv_1)\rangle}{\langle L
(\rv_1)\rangle} - \int d\rv_2 \left[ \frac{\langle
L(\rv_1)L(\rv_2)\rangle-\langle L(\rv_1)\rangle\langle
L(\rv_2)\rangle}{\langle L(\rv_1)\rangle\langle L(\rv_2)\rangle}
\right]\Hv(\rv_1,\rv_2)\langle\Pv(\rv_2)\rangle - \label{master}
\end{equation}
\begin{equation}
\int d\rv_2 d\rv_3 \left[ \frac{\langle
L(\rv_1)L(\rv_2)L(\rv_3)\rangle}{\langle L(\rv_1)\rangle\langle
L(\rv_2)\rangle}-\frac{\langle L(\rv_1)L(\rv_2)\rangle\langle
L(\rv_2)L(\rv_3)\rangle}{\langle L(\rv_1)\rangle\langle
L(\rv_2)\rangle\langle
L(\rv_3)\rangle}\right]\Hv(\rv_1,\rv_2)\Hv(\rv_2,\rv_3)\langle\Pv(\rv_3)\rangle-...
\end{equation}
From this expression and Eq. (\ref{homogenization}), we obtain the
following exact expansions for statistically homogeneous media involving the effective dielectric
tensor:
\begin{equation}
\beta_{pq}^2\phi_p^2({\bf \sigma}_{e} -\sigma_q{\bf I})^{-1}({\bf
\sigma}_{e} +2\sigma_q{\bf I})=\phi_p\beta_{pq}{\bf I} -
\sum_{n=2}^{\infty}{\bf A}^{(p)}_n\beta_{pq}^n, (p\neq q),\label{fullsolution}
\end{equation}
where $p\neq q$,
\begin{equation}
\beta_{pq}=\frac{\sigma_p-\sigma_q}{\sigma_p+(d-1)\sigma_q} =
\frac{\varepsilon_p-\varepsilon_q}{\varepsilon_p+(d-1)\varepsilon_q},\label{beta}
\end{equation}
and the $n$-point tensor coefficients ${\bf A}^{(p)}_n$ are certain integrals
over the $n$-point correlation functions $S_n^{(p)}$ associated with phase $p$. In particular, for $n=2$
\begin{equation}
{\bf A}^{(p)}_2=\frac{d}{\Omega}\int d\rv
\;\tv^{(p)}(\rv)\left[S^{(p)}_2
(\rv)-\phi_p^2\right],\label{two-point-integral}
\end{equation}
where 
\begin{equation}
\tv^{(p)}(\rv)=\Omega k_q^2 \Hv (\rv), 
\end{equation}
and $\Omega$ is the solid angle of a sphere in
dimension $d$, and $\Hv(\rv)$ is given by Eq. (\ref{H-definition}).
For any $n >2$ , we have 
\begin{equation}
{\bf A}^{(p)}_n=\left(
\frac{-1}{\phi_p}\right)^{n-2}\left(\frac{d}{\Omega}\right)^{n-1}\int
d\rv_2 ... \int d\rv_n
\tv^{(p)}(\rv_1,\rv_2)\cdot\tv^{(p)}(\rv_2,\rv_3)\cdot\cdot\cdot
\tv^{(p)}(\rv_{n-1},\rv_{n})\Delta^{(p)}_n(\rv_1,...,\rv_n),
\label{n-point-integral}
\end{equation}
where  $\Delta^{(p)}_n(\rv_1,...,\rv_n)$ the determinant is given by
\begin{equation}
\Delta^{(p)}_n =  \left|
\begin{array}{llll}
 S_2^{(p)}(\rv_1,\rv_2) & S_1^{(p)}(\rv_2) & \cdots & 0 \\
 S^{(p)}_3(\rv_1,\rv_2,\rv_3) & S^{(p)}_2(\rv_2,\rv_3) & \cdots & 0 \\
 \vdots & \vdots & \ddots & \vdots \\ S^{(p)}_n(\rv_1,...,\rv_n) &
 S^{(p)}_{n-1}(\rv_2,...,\rv_n) & \cdots & S^{(p)}_2(\rv_{n-1},\rv_n)
\end{array}
\right|
\label{det}
\end{equation}
and 
\begin{equation}
S^{(p)}_n(\rv_1,\rv_2,\ldots,\rv_n)\equiv\langle I^{(p)}(\rv_1)I^{(p)}(\rv_2)\cdots I^{(p)}(\rv_n)\rangle .\label{correlationfunction}
\end{equation}
is the  $n$-point correlation function of phase $p$. For statistically homogeneous media (the case for which this formulation applies),
the latter quantity is translationally invariant and therefore
depends only on relative displacements, i.e., $S^{(p)}_n(\rv_1,\rv_2,\dots,\rv_n)=S^{(p)}_n(\rv_{12},\rv_{13},\ldots,\rv_{1n})$,
where we have chosen point 1 to be the origin, and, in particular, $S^{(p)}_1(\rv_1)=\phi_p$.

\noindent{\it Remarks}

\noindent{1. We see that the 
integral-equation approach given here is entirely general.
It is equally well suited to any problem in which 
the local and averaged constitutive relations have the same form,
and thus the appropriate dyadic Green's function would replace
Eq. (\ref{GF}). Indeed, the determinant (\ref{det}) 
for the dynamic problem considered here is exactly the
same as that in the static problem, as given originally in
Ref. \onlinecite{Torquato1}.}
\smallskip  

\noindent{2. Note that Eq. (\ref{fullsolution}) represents two different series
expansions: one  for $q=1$ and $p=2$ and the other for $q=2$ and $p=1$.}
\smallskip

\noindent{3. The quantity $\beta_{pq}$, given in Eq. (\ref{beta}), is
the strong-contrast expansion parameter, the form of which is a direct
consequence of the choice of exclusion volume associated with the
source-term of the Green's function.  Any shape besides the sphere
would have necessarily led to a different expansion parameter and
therefore to significantly different convergence properties \cite{RHM}.  Clearly, $\beta_{pq}$ may lie within the range
$-(d-1)^{-1}\leq \beta_{pq} \leq 1$.  The radius of convergence of the
series given in Eq. (\ref{fullsolution}) is thus greatly widened
beyond that of a weak-contrast expansion (i.e., the simple difference of the
phase dielectric constants).}

\noindent{4. In the purely static case, the series represented by
(\ref{fullsolution}) may be regarded as expansions that perturb
around the optimal structures that realize the generalized
Hashin-Shtrikman bounds \cite{HS} derived by Willis \cite{Willis}, as discussed by
Torquato \cite{RHM}.  In particular, these optimal structures are
certain dispersions in which there is a disconnected, dispersed phase
in a connected matrix phase.  The lower bound corresponds to the case
when the high-dielectric-constant phase is the dispersed, disconnected
phase and the upper bound corresponds to the instance in which the
high-dielectric constant phase is the connected matrix.  Thus, we
expect that for the dynamic problem under consideration, the
first few terms of the expansion (\ref{fullsolution}) with $q=1$ and $p=2$ will yield a
reasonable approximation of ${\bf \varepsilon_e}$ for two-phase
media, depending on whether the high-dielectric phase
is below or above its percolation threshold, as discussed
further in Section II.C.}

\subsection{Macroscopically Isotropic Media}

Consider, as we will to a large extent in this paper, two-phase media that are
statistically homogeneous and isotropic, and thus macroscopically isotropic.  In this
case, the effective dielectric tensor is proportional to the identity
tensor, and can thus be treated as a simple scalar, namely, the
effective dielectric constant of the medium.  This of course
simplifies the calculation enormously.  We thus take the trace of both
sides of Eq. (\ref{fullsolution}) and divide by $d$.  Thus, the full
expression in the isotropic case is given by
\begin{equation}  
\beta_{pq}^2\phi_p^2\beta^{-1}_{eq}=\phi_p\beta_{pq} -
\sum_{n=2}^{\infty}A^{(p)}_n\beta_{pq}^n,\label{fullsolutionisotropic}
\end{equation}
where $A^{(p)}_n=\Tr \left[{\bf A}^{(p)}_n\right]/d$ and $\beta_{eq}$
is the effective polarizability, given by
\begin{equation}
\beta_{eq}=\frac{\varepsilon_e-\varepsilon_q}{\varepsilon_p+(d-1)\varepsilon_q}.
\end{equation}
The scalar two-point coefficient, as specified by taking the trace of (\ref{two-point-integral}), is given by
\begin{equation}
A^{(p)}_2=\frac{d}{\Omega}\int d\rv
\frac{\Tr\left[\tv^{(p)}(\rv)\right]}{d}\left[S^{(p)}_2
(\rv)-\phi_p^2\right]=2k_q^2\int_0^\infty dr \exp(i k_q
r)r\left[S^{(p)}_2 (r)-\phi_p^2\right],\label{scalarA2}
\end{equation}
with this integral being straightforwardly carried out either
numerically or analytically, depending on the form of the correlation
function. Provided that the correlation function $S^{(p)}_2 (r)$ decay
sufficiently rapidly to its long-range value of $\phi_p^2$, the integral
in (\ref{scalarA2}) will be convergent.

Note that by Eq. (\ref{scalarA2}), $A^{(p)}_2$ must be zero in the
purely static problem ($\omega=0$) if the medium is statistically
homogeneous and isotropic.  Thus, in the static problem, assuming statistical 
homogeneity and isotropy, two-point information is actually incorporated
even though $A^{(p)}_2$ is zero. This subtle point is elaborated 
in Ref. \onlinecite{RHM}. This suggests that our formalism is extremely well suited to
the non-static problem in the low frequency limit being considered
here.  Expanding the two-point coefficient  $A^{(p)}_2$ given
in Eq. (\ref{scalarA2}) through third order in $k_q$ about $k_q=0$ yields
\begin{equation}
A^{(p)}_2= 2k_q^2\int_0^\infty dr r\left[S^{(p)}_2
(r)-\phi_p^2\right]+ 2 i k_q^3\int_0^\infty dr r^2\left[S^{(p)}_2
(r)-\phi_p^2\right] + {\cal O}(k_q^4).\label{k-expansion}
\end{equation}
It should be expected that the imaginary component of the effective
dielectric constant of statistically homogeneous and isotropic systems should be
positive; otherwise the homogenized coherent wave would be amplified
rather than attenuated.  A simple analysis of Eq. (\ref{k-expansion})
bears this out.  The second term on the right hand side, i.e., the
leading-order imaginary component, is nothing more than the
zero-wave-vector structure factor of the composite material.  Since
$S^{(p)}_2 (r)$ is obtained from a realizable configuration, this is
necessarily positive \cite{TorquatoRealizable}.  This is true because
the structure factor is nothing but the squared norm of the Fourier
transform of the function $\Ind^{(p)}(\rv)-\phi_p$.  The leading-order
term of the imaginary component in the expansion given in
Eq. (\ref{k-expansion}) is directly proportional to the volume
integral of the function $S^{(p)}_2(r)-\phi_p^2$.  This is exactly
$(\phi_2 C_{\infty})^2v_0$, where $C_{\infty}$ is the coarseness of
the composite structure in the limit of an infinitely large
window, and $v_0$ is the window volume \cite{RHM,LuTorquato}.  Thus,
the value of the nonnegative imaginary part of the effective
dielectric constant to leading order is determined by
local-volume-fraction fluctuation over very large windows, which may
be taken to be a crude measure of disorder in the system.
\cite{StillingerTorquato}  Note also that $C_\infty$ is proportional
to the single-scattering intensity of scattered radiation from the
random medium in the infinite-wavelength limit \cite{RHM,DebyeAnderson}.

Another important fact that emerges from this formalism is that at the
two-point level the correction to the static effective dielectric
constant only enters at second order in $k_q$ (i.e., second order in the
frequency).  It is hence a relatively small contribution.  The
imaginary term enters only at third order in the wave number.

Now, let us consider the simplification of the three-point coefficient
$A^{(p)}_3$.  From the general expression (\ref{n-point-integral}), we find
\begin{equation}
A^{(p)}_3=\frac{9}{(4\pi)^2}\int d^3{\bf r}_{12}\int d^3{\bf r}_{13}
\tv^{(p)}(\rv_{12})\cdot\tv^{(p)}(\rv_{12})\left[S^{(p)}_3(\rv_{12},\rv_{13})
-\frac{S^{(p)}_2(\rv_{12})
S^{(p)}_2(\rv_{13})}{\phi_p}\right]\label{three-point}
\end{equation}
The first few terms of the Taylor expansion of (\ref{three-point}) about $k_q=0$
is given by
\begin{eqnarray}
A^{(p)}_3&=& \frac{18}{(4\pi)^2}\int\frac{d^3{\bf
r}_{12}}{r_{12}^3}\int\frac{d^3{\bf
r}_{13}}{r_{13}^3}\left[1+\frac{k_q^2}{6}(r_{12}^2+r_{13}^2)\right]
P_2(\mu)\left[S^{(p)}_3(\rv_{12},\rv_{13})
-\frac{S^{(p)}_2(\rv_{12}) S^{(p)}_2(\rv_{13})}{\phi_p}\right] + {\cal O}(k_q^4) \nonumber\\
&=&9\int_0^\infty\frac{dr}{r}\int_0^\infty\frac{ds}{s}\int_{-1}^{1}
d\mu\left[1+\frac{k_q^2}{6}(r^2+s^2)\right]
P_2(\mu)\left[S^{(p)}_3(r,s,\mu)-\frac{S^{(p)}_2(r)
S^{(p)}_2(s)}{\phi_p}\right]+ {\cal O}(k_q^4)\label{k-expansion-3pt}.
\end{eqnarray}
where $\mu=\cos\theta$, $\theta$ is the angle between $\rv_{12}$ and $\rv_{13}$ and
$P_2$ is the second-order Legendre polynomial.  The simplified second line
is obtained by exploiting the homogeneity and isotropy of
the medium, reducing the original  six-dimensional integral to a
three-dimensional one \cite{Torquato3point1}.  Here 
 $r=|\rv_{12}|$ and $s=|\rv_{13}|$ are the side lengths
of a triangle and $\theta$ is the angle between these sides.  Note that the
third-order term in $k_q$ is exactly zero, the next-lowest-order contribution
to the real component must be of the order of $k_q^4$ (or higher), and
the next-lowest-order contribution to the imaginary component must be
of the order of $k_q^5$ (or higher).

In Section IV of this paper, estimates for the effective dielectric constant  that include three-point
information will be presented for the fully-penetrable-sphere model.

\subsection{Two- and Three-Point Approximations}

Practically speaking, it is difficult to ascertain four-point and
higher-order correlation functions which therefore prohibits an exact
evaluation of the expansion for the effective dielectric tensor in
Eq. (\ref{fullsolution}).  However, as shown in the static problem \cite{Torquato1,RHM}, 
lower-order truncations of the exact expansion
(\ref{fullsolution}) at the two-point and three-point levels have
proved to be accurate approximations for the effective dielectric
tensor.  For the macroscopically isotropic case, the two-point and
three-point approximations used to calculate the effective dielectric
constant, obtained by truncating the series in
Eq. (\ref{fullsolutionisotropic}), are given by
\begin{equation}
\left[\frac{\varepsilon_p-\varepsilon_q}{\varepsilon_p+(d-1)\varepsilon_q}\right]\phi_p^2\left[\frac{\varepsilon_e-\varepsilon_q}{\varepsilon_e+(d-1)\varepsilon_q}\right]^{-1}=\phi_p
-
A^{(p)}_2\left[\frac{\varepsilon_p-\varepsilon_q}{\varepsilon_p+(d-1)\varepsilon_q}\right],
(p\neq q)\label{two-point-isotropic},
\end{equation}  
and
\begin{equation}
\left[\frac{\varepsilon_p-\varepsilon_q}{\varepsilon_p+(d-1)\varepsilon_q}\right]\phi_p^2\left[\frac{\varepsilon_e-\varepsilon_q}{\varepsilon_e+(d-1)\varepsilon_q}\right]^{-1}=\phi_p
-
A^{(p)}_2\left[\frac{\varepsilon_p-\varepsilon_q}{\varepsilon_p+(d-1)\varepsilon_q}\right]
-
A^{(p)}_3\left[\frac{\varepsilon_p-\varepsilon_q}{\varepsilon_p+(d-1)\varepsilon_q}\right]^2,
(p\neq q)\label{three-point-isotropic},
\end{equation}
respectively, where $A^{(p)}_2$ is given by Eq. (\ref{k-expansion}),
and $A^{(p)}_3$ is given by Eq. (\ref{k-expansion-3pt}).  The
two-point approximation (\ref{two-point-isotropic}) is exact to second
order in $\varepsilon_p-\varepsilon_q$ for any $\phi_p$ and is exact
to first order in $\phi_p$ for any phase-contrast ratio
$\varepsilon_p/\varepsilon_q$.  The three-point approximation
(\ref{three-point-isotropic}) is exact to third order in
$\varepsilon_p-\varepsilon_q$ for any $\phi_p$ and is exact to first
order in $\phi_p$ for any phase-contrast ratio
$\varepsilon_p/\varepsilon_q$.  However, the most difficult cases to
treat theoretically are when both the volume fraction and
phase-contrast ratio are significantly different from zero and unity,
respectively.  The strong-contrast approximations
(\ref{two-point-isotropic}) and (\ref{three-point-isotropic}) are
expected to provide reasonable estimates of $\varepsilon_e$ for 
a certain class of dispersions in this
more difficult regime because they are perturbations of the
strong-contrast expansions in the infinite-wavelength limit (pure
static case), which have been shown to be in excellent agreement
with both precise computer-simulation and experimental data
for a variety of dispersions \cite{Torquato1, RHM}.
Specifically, if $\epsilon_p > \epsilon_q$,
(\ref{two-point-isotropic}) and (\ref{three-point-isotropic}) with
$q=1$ and $p=2$ will yield good estimates of $\varepsilon_e$
provided that phase 2 is below its percolation threshold and that the
typical cluster size of phase 2 is sufficiently small \cite{Torquato1,RHM}.  On the other hand, if phase 2 is above its
percolation threshold, (\ref{two-point-isotropic}) and
(\ref{three-point-isotropic}) with $q=2$ and $p=1$ should provide good
estimates of $\varepsilon_e$.  

\section{Model Microstructures}

\begin{figure}[h]
\begin{center}
\includegraphics[width=5in]{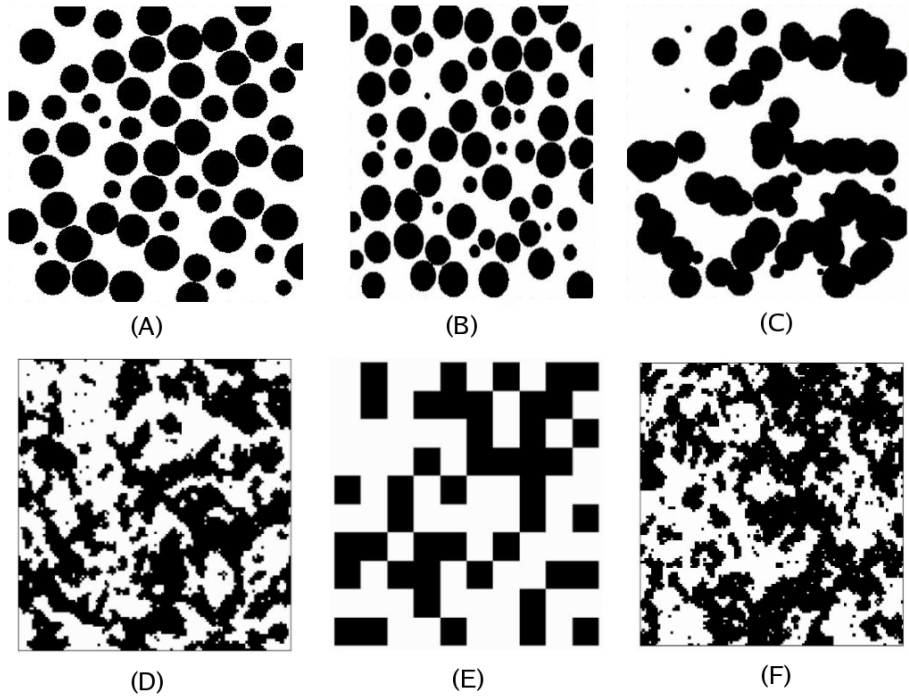}
\caption{(Color online) Two-dimensional slices of the
three-dimensional microstructures described in this section at volume
fraction $\phi_2=0.5$.  They are hard spheres (A), hard oriented spheroids (B),
fully penetrable spheres (C), Debye random media \cite{YeongTorquato}
(D), random checkerboard (E) (in any plane perpendicular to a
principal axis), and power-law-correlated materials \cite{YeongTorquato} (F).
Both the fully-penetrable-sphere model and random checkerboard at
$\phi_2=0.5$ are above their respective percolation thresholds for the
black phase 2, even though planar cuts through these samples do not
reveal that the black phase is indeed percolating in three dimensions.
We expect that the Debye random medium and power-law-correlated
materials shown here also percolate at $\phi_2=0.5$.
}\label{fig:microstructures}
\end{center}
\end{figure}

\begin{figure}[h]
\begin{center}
\includegraphics[width=4in]{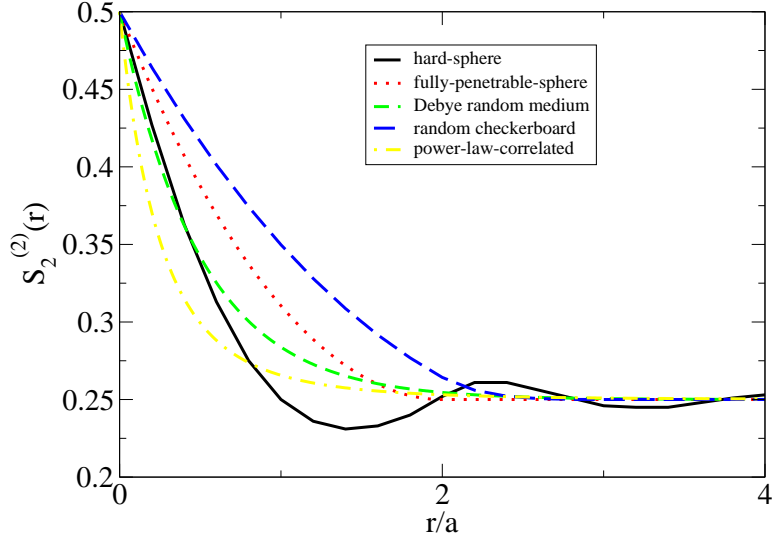}
\caption{(Color online) Plots of the two-point correlation
function $S^{(p)}_2(r)$ for the five isotropic models detailed in
this section: hard spheres, fully penetrable spheres, Debye random
medium, random checkerboard, and a power-law-correlated material.  The
volume fraction for each model is $\phi_2=0.5$.  Note that the
correlation lengths for each model are roughly equal to one another.}
\label{fig:S2s}
\end{center}
\end{figure}

We study a number of model microstructures and their corresponding
two-point correlation functions.  In particular, we examine dispersions
of hard spheres, hard oriented spheroids and fully penetrable spheres
as well as Debye random media, the random checkerboard, and 
power-law-correlated materials.  Except for the hard-ellipsoid
model, all of the other models are statistically homogeneous
and isotropic. We take $a$ to
be a characteristic length scale for each model.  These
microstructures, depicted in Fig. \ref{fig:microstructures} are
explicitly described in this section, and  we take $q=1$
and $p=2$.  The two-point correlation functions for each of these
models is depicted in Fig. \ref{fig:S2s}, at volume fraction
$\phi_2=0.5$, with correlation lengths roughly equal to one another.
The ``correlation length'' $l_c$ is the range $\left[0,l_c\right]$ in
$r$ over which the magnitude of $S_2^{(2)}-\phi_2^2$ is appreciably fluctuating  
around zero \cite{RHM}.

\subsection{Equilibrium Hard Spheres} 
Here we consider the well-known equilibrium distribution of hard
spheres of radius $a$ (phase 2) in a matrix (phase 1) \cite{HansenMcDonald}.  For this model, depicted in
Fig. \ref{fig:microstructures}(A), all non-overlapping configurations
are equally probable.  This equilibrium model is athermal, i.e., the behavior of
the system is temperature-independent.  The pair correlation function $g_2(r)$
for particle centers may be expressed in closed form in
Fourier space using the Percus-Yevick approximation, as described in
Ref. \onlinecite{RHM}.  The two-point function $S^{(2)}_2(r)$ can be obtained from the pair
correlation function via the following equation \cite{TorquatoStell1985}:
\begin{equation}
S^{(2)}_2(r)=\rho v_2^{(int)}(r;a) + \rho^2 g_2 \otimes m \otimes m,\label{hs}
\end{equation}
where $\rho$ is the number density of spheres, $v_2^{(int)}(r;a)$ is the intersection volume
between two spheres of radius $a$ that are a distance $r$ apart, which
is given by
\begin{equation}
v_2^{(int)}(r;a)=\left\{ \begin{array}{l} 1 - \frac{\displaystyle 3}{\displaystyle 4}\frac{\displaystyle r}{\displaystyle a} +
\frac{\displaystyle 1}{\displaystyle 16}\left(\frac{\displaystyle r}{\displaystyle a}\right)^3, \mbox{ if } r < 2a \\ 0,
\mbox{ otherwise}\\
\end{array}\right . ,
\end{equation} 
the symbol $\otimes$ denotes the convolution of two functions, and the
step function $m$ is defined by
\begin{equation}
m(\rv) = \left\{ \begin{array}{l}
1, \mbox{ if } |\rv| < a \\
0,  \mbox{ otherwise}\\ 
\end{array}\right . .
\end{equation}
Note that the volume fraction of spheres is given by $\phi_2=\rho 4\pi a^3/3$.
The two-point correlation function $S^{(2)}_2(r)$ is plotted in
Fig. \ref{fig:S2s}.  The most convenient method of obtaining
$S^{(2)}_2(r)$ via Eq. (\ref{hs}) is via Fourier transform techniques.
The reason for this is that in Fourier space, the
convolution operators become simple products, and the Percus-Yevick
pair correlation function may be expressed analytically.  In order to
obtain $S_2^{(2)}(r)$ in real space, numerical inversion from
Fourier representation must be performed.

Of all the isotropic disordered models considered here, the hard-sphere
system may be thought to possess the greatest degree of order because
its coarseness $C_\infty$ is minimized among the structures
considered.  For example, consider volume fraction $\phi_2=0.5$, at
which spheres are slightly above their freezing volume fraction.  The
correlation function $S^{(2)}_2(r)-\phi_2^2$ exhibits oscillation
above and below zero, suggesting strong short-range correlations and
anti-correlation among the spheres in the system.  Calculations show
that as a result of this property, the imaginary component of the
dielectric constant is very small compared with that of other model microstructures
considered here.

\subsection{Equilibrium Hard Spheroids} 
We also consider an equilibrium dispersion of hard spheroids
(phase 2), or ellipsoids of revolution, in a matrix (phase 1), which
are constrained to have the same orientation (see
Fig. \ref{fig:microstructures}(B)) along the $z$ axis. Because
there is an axis of symmetry, spheroids possess azimuthal symmetry. 
This statistically homogeneous but anisotropic dispersion is considered here in
order to demonstrate the application of our formalism to
macroscopically anisotropic media, i.e., materials with an effective
dielectric tensor with non-equal diagonal terms in the principal axes frame.   
The spheroid shape is defined by
\begin{equation}
(x^2+y^2)/a^2 + z^2/b^2 = 1,\label{spheroid}
\end{equation}
where $a$ and $b$ are the semi-axes of the spheroid with $b$ being
along the axis of symmetry (i.e., $z$ axis).  We define the aspect ratio as
$b/a$ so that $b/a>1$ and $b/a<1$ corresponds to prolate
and oblate spheroids, respectively. In a previous work, Torquato and Lado \cite{TorquatoLado} showed that the correlation function for 
any dispersion of oriented spheroids (in equilibrium or not)
at number density $\rho$ could be transformed directly from the hard-sphere correlation
function at the same number density by an affine (linear) transformation
of the coordinate system. Taking phase 2 to be the spheroid phase, this transform is
defined by
\begin{equation}
S_{2,HS}^{(2)}(\rv;b/a) =
S_{2,HS}^{(2)}\left[\sigma_0(r_{12}/\sigma(\theta));1\right],\label{spheroidS2}
\end{equation}
where
\begin{equation}
\sigma(\theta) =
\frac{2a}{\left[1-(1-a^2/b^2)\cos^2(\theta)\right]^{1/2}},
\end{equation}
where $\sigma_0=2a$, $\theta$ is the polar angle between the $z$ axis and the radial
vector $\rv$, and $S_{2,HS}^{(2)}(\rv;1)$
is the hard-sphere correlation function at the same
number density. Here we employ the equilibrium hard-sphere model
to get the corresponding expression for the equilibrium
hard-oriented-spheroid system.

\subsection{Fully Penetrable Spheres} 
A fully-penetrable-sphere model is composed of spheres of radius
$a$ (phase 2) whose centers are completely uncorrelated in space \cite{RHM} (see Fig. \ref{fig:microstructures}(C)).  For this dispersion,
the particle phase percolates at $\phi_2=0.2895\pm 0.0005$ \cite{RintoulTorquato} (i.e., the medium contains infinite clusters of
phase 2), and the matrix phase (phase 1) percolates until a volume
fraction of spheres of $97\%$.  The $n$-point correlation function for
this system, for phase 1, is given in Ref. \onlinecite{RHM} as
\begin{equation}
S^{(1)}_n(\rv^n) = \exp\left[-\rho v_n (\rv^n;a)\right],
\end{equation}
where the number density $\rho$ is defined here by
$\phi_1=\exp(-\eta)$, with $\eta=\rho 4\pi a^3/3$, and the
function $v_n (\rv^n;a)$ gives the union volume of $n$ spheres of
radius $a$ with centers defined by the position coordinates
$\rv^n\equiv \rv_1,\rv_2,\ldots,\rv_n$. Thus, we may write, for the
particle phase,
\begin{equation}
S^{(2)}_2(r) = 1 - 2\phi_2 + \exp\left[ -\eta \frac{v_2(r;a)}{v_1(a)}
\right],
\end{equation}  
where we have
\begin{equation}
\frac{v_2(r;a)}{v_1(a)}=2\Theta (r-2a) + \left[ 1 +
\frac{3}{4}\frac{r}{a} - \frac{1}{16}\left(\frac{r}{a}\right)^3
\right]\Theta(2a-r),\label{fpsunion}
\end{equation}
and $\Theta(x)$ is the unit step function.  The correlation
function $S^{(2)}_2(r)$ is plotted in Fig. \ref{fig:S2s}.  Here, we
have used the fact that we can relate the union and intersection
volume of two spheres by the equation
\begin{equation}
v_2(r;a)=2v_1(a) - v_2^{(int)}(r;a).
\end{equation}    
The function $S^{(2)}_3$ may be obtained in similar way.  It involves
an expression for the intersection volume of three spheres, given
originally by Powell in Ref. \onlinecite{Powell}, and discussed in detail in
Ref. \onlinecite{RHM}.  In the next section, we present estimates of the
effective dielectric constant as obtained from two- and three-point
approximations for this model. Compared with the hard-sphere model,
the fully-penetrable-sphere dispersion is significantly more
disordered.  This is reflected in the appreciably large difference
between the imaginary components of their effective dielectric
constants, as will be described.

\subsection{Debye Random Medium} 
The two-point correlation function for a
Debye random medium is given by
\begin{equation}
S^{(2)}_2(r)-\phi_2^2=\phi_2(1-\phi_2)\exp(-r/\gamma), \label{drm}
\end{equation}
with $\gamma>0$.  This correlation function is plotted in
Fig. \ref{fig:S2s}.  In the following analysis, we take $\gamma=a/2$.
This correlation function, first proposed by Debye \cite{DebyeAnderson}, was imagined to correspond to porous media with
cavities of random shapes and sizes (a realization of this medium is
depicted in Fig. \ref{fig:microstructures}(D)).  However, it was not
known until recently that such a correlation function was indeed
realizable by a two-phase medium.  Using a ``construction'' procedure,
Yeong and Torquato \cite{YeongTorquato} demonstrated that the
correlation function defined in Eq. (\ref{drm}) is realizable and
dubbed such systems Debye random media (see also 
Refs. \onlinecite{RHM} and \cite{TorquatoRealizable} for further discussion of the realizability
of Debye random media).  Note that if the two component dielectrics
and their respective volume fractions are interchanged, the
correlation function given in Eq. (\ref{drm}) remains invariant.  This
property is called {\it phase-inversion symmetry}.  A composite is
phase-inversion symmetric if the morphology of phase 1 at volume
fraction $\phi_1$ is statistically identical to that of phase 2 at the
same volume fraction, $\phi_1$ \cite{YeongTorquato,RHM}.  None of the
sphere and spheroid dispersions possess this property.  It is also
important to note that the percolation behavior of Debye random media
has not been investigated to date.  However, based on our knowledge of
the percolation threshold of the fully-penetrable-sphere model and
visual inspection of the planar cut through a three-dimensional Debye
random medium shown in Fig. \ref{fig:microstructures}, we expect that
the percolation threshold of the latter is substantially below
$\phi_2=0.5$.

\subsection{Random Checkerboard} 
The random checkerboard model is produced by partitioning space into
cubes such that each cube is assigned phase 1 or phase 2 randomly
according to the volume fraction (see
Fig. \ref{fig:microstructures}(E)).  The cubes have side length
$D=2a$.  The calculation of the radially averaged two-point
correlation function for this model is not presented here.
A detailed derivation of this result may be found in Ref. \onlinecite{RHM}.
The radially averaged two-point function, $S^{(2)}_2(r)$, is plotted in
Fig. \ref{fig:S2s}.  This model may be thought of as being more
``ordered' than, for example, the Debye random medium because the phases
are confined to lie on a grid.  In our calculations, presented in the
next section, we find that the random checkerboard has a greater
imaginary component, however, than that of hard spheres at volume
fraction $\phi_2=0.5$.  This makes intuitive sense: in this regime,
the hard-sphere system is near crystallization; there is no such
order-disorder transition for the random checkerboard model.  As in
the Debye random medium, this model possesses phase-inversion
symmetry.  For site percolation on a simple cubic lattice with
nearest-neighbor connectivity criteria, phase 2 would percolate at a
volume fraction of $\phi_2=0.312$.  However, for the related
dielectric and conductivity problems, it is known that edges and corner points
will contribute to the effective properties and therefore nearest,
next-nearest, and next-next-nearest connectivity criteria must be used.  This will result in
a substantially lower percolation threshold than $\phi_2=0.312$, as
one can ascertain from the analogous two-dimensional percolation
problem on a square lattice.  In summary, the percolation threshold for
the three-dimensional random checkerboard should be considerably lower
than that for fully penetrable spheres ($\phi_2=0.2895$).

\subsection{Power-Law-Correlated Materials} 
The two-point correlation function for these
composites are given by
\begin{equation}
S^{(2)}_2(r)-\phi_2^2=\frac{\phi_2(1-\phi_2)a^n}{(r+a)^n},\label{power-law}
\end{equation}
where the exponent $n$ must obey the inequality $n\ge 3$ for 
the Fourier transform of the left side of (\ref{power-law}) to exist.
The two-point correlation function $S^{(2)}_2$ is plotted in Fig. (\ref{fig:S2s})
with $n=4$, which is the value of the exponent that will be considered 
throughout the rest of the paper. A single realization
of a power-law-correlated material with this correlation function is depicted in
Fig. \ref{fig:microstructures}(F).  The degree of correlation in this
model is of course highly dependent upon the exponent $n$.  Note that as in the
cases of the Debye random medium and the random checkerboard, the
power-law-correlated material is phase-inversion symmetric.  This
correlation function is being put forward in this study for the first
time.  However, it is not necessarily true that any proposed
functional form of $S_2$ is physically realizable.  Any correlation
function that corresponds to a realizable material must satisfy a
number of necessary conditions \cite{TorquatoRealizable}.  Namely, (1)
$0\leq S^{(2)}_2(\rv)\leq\phi_2$, (2) the radial derivative of
$S^{(2)}_2$ must be strictly negative at the origin, (3) it must obey
the triangle inequality $S^{(2)}_2(\rv)\geq S^{(2)}_2({\bf
s})+S^{(2)}_2({\bf t})-\phi_2$, where $\rv = {\bf t} - {\bf s}$, and
(4) the Fourier transform of $S^{(2)}_2(\rv)-\phi_2^2$ must be
everywhere non-negative.  We have tested each of these conditions for
the power-law correlation function, and they are all satisfied.  For
all other models considered in this work have either been known
to be realizable or have been recently shown to be \cite{YeongTorquato,RHM}.  We
choose to study power-law-correlated materials because they are
reminiscent of structures that are {\it scale-free}.  While the
correlation function given in Eq. (\ref{power-law}) is, strictly speaking, not scale-free
(i.e., a function directly proportional to $1/r^n$), it does
asymptotically approach this behavior for $r\gg a$.  A purely
scale-free function does not satisfy the known realizability constraints.
Model microstructures that have scale-free correlation functions exhibit interesting
clustering and percolation properties \cite{RHM}.  This model's percolation behavior will be
studied in greater detail in a forthcoming study \cite{Jiao}.   As
in the case of the Debye random medium, percolation behavior of
power-law-correlated materials has not been investigated to date.
However, for the same reasons given in Section III.D, we expect that
the percolation threshold of the latter is substantially below
$\phi_2=0.5$.

\section{Results}

In this section, we present results  for the effective dielectric
constant as predicted by the two-point approximation
(\ref{two-point-isotropic}) for each of the model microstructures discussed in the
previous section, i.e., hard spheres, hard oriented spheroids, fully penetrable
spheres, Debye random media, random checkerboard, and
power-law-correlated materials (with exponent $n=4$).  Henceforth, we take
$\varepsilon_2\geq \varepsilon_1$, and assume both $\varepsilon_1$ and
$\varepsilon_2$ are real.  We take the wave number for a wave
propagating through phase 1 to be $2\pi/(60a)$.  If phase 1 has
dielectric constant 1, then this corresponds to a propagation
frequency of 5 GHz.   For reasons given in Section II.C, we take $q=1$ and
$p=2$ if phase 2 is below its percolation threshold, and we take $p=1$
and $q=2$ if it is above its percolation threshold.  In the case of the
fully-penetrable-sphere model, we also evaluate the three-point
approximation (\ref{three-point-isotropic}).

\subsection{Two-Point Estimates}

\subsubsection{Isotropic Media}

For the hard-sphere model, we employ the two-point approximation
given in Eq. (\ref{two-point-isotropic}) with $q=1$ and $p=2$ where
the spheres comprise phase $2$, for $\phi_2\leq 0.5$ (see discussion
in Section II.C).  This choice of two-point approximation is expected
to be accurate because the equilibrium hard-sphere model percolates at ``jammed"
states, which are substantially higher than $\phi_2=0.5$ \cite{RHM}.
In the case of the fully-penetrable-sphere model, we plot our
results for a fixed phase contrast ratio but over the entire volume
fraction range because in this case we know its nontrivial percolation
threshold (see Section III.C). For this model, we expect
(\ref{two-point-isotropic}) with $p=2$ and $q=1$ to be valid for
volume fractions well below its percolation threshold of
$\phi_2=0.2895$, and then with $p=1$ and $q=2$ to be valid well above
it.  In the intermediate region, which is taken to be $0.2 < \phi_2 <
0.4$, we interpolate between these two regimes using a spline fit in
order to approximately account for the fact that this medium is
crossing its percolation threshold.   For Debye random media, random
checkerboard, and power-law-correlated materials (in which we take 
the exponent $n=4$), the percolation
thresholds are not known.  Therefore, we limit ourselves to analyzing
the effective dielectric constant of these models for volume
fractions that are assumed to be below these thresholds (e.g.,
$\phi_2=0.1$), and well above them (e.g., $\phi_2=0.5$).  In the former
case, we use the two-point approximation (\ref{two-point-isotropic})
with $p=2$ and $q=1$, and in the latter case with $p=1$ and $q=2$.
For all of the five isotropic models, we employ the two-point estimate
(\ref{two-point-isotropic}) with the expansion given in
Eq. (\ref{k-expansion}) through third order in $k_q$, and choose the
correlation lengths to be roughly equal to one another.  Different
two-point approximations are employed for the various statistically homogeneous and isotropic
model microstructures below and above their percolation thresholds for reasons given
at the end of Section II.C.

For the disordered models described in Section III, we
present in Table \ref{Table:S2} the two lowest-order coefficients of
$k_q a$ in the expansion of $A^{(p)}_2$, as given in
Eq. (\ref{k-expansion}).  Note that the zeroth-order term is zero,
because $A^{(p)}_2$ is zero in the isotropic static problem.  As can
be seen from Eq. (\ref{scalarA2}), there can be no first-order term
either.  The second-order term is necessarily real, and the
third-order term necessarily imaginary.  We call the $j^{th}$ order
coefficient in this expansion $\alpha^{(2)}_j$, such that
$A^{(p)}_2=\sum_{j=2}^{\infty}\alpha^{(2)}_j (k_q a)^j$.

\begin{table}
\begin{tabular}{|c||c|c||c|c||c|c||c|c||c|c|}
\hline
 &\multicolumn{2}{c|}{HS}&\multicolumn{2}{c|}{FPS}&\multicolumn{2}{c|}{DRM}&\multicolumn{2}{c|}{RC}&\multicolumn{2}{c|}{PLC}\\
 \cline{2-11}
 &$\alpha^{(2)}_2$&$\alpha^{(2)}_3$&$\alpha^{(2)}_2$&$\alpha^{(2)}_3$&$\alpha^{(2)}_2$&$\alpha^{(2)}_3$&$\alpha^{(2)}_2$&$\alpha^{(2)}_3$&$\alpha^{(2)}_2$&$\alpha^{(2)}_3$\\
 \hline\hline
 $\phi_2=0.1$&0.0512&0.0304$i$&0.0700&0.0579$i$&0.0450&0.0450$i$&0.108&0.116$i$&0.0300&0.0600$i$\\
 $\phi_2=0.2$&0.0658&0.0279$i$&0.120&0.0990$i$&0.0800&0.0800$i$&0.193&0.206$i$&0.0533&0.1067$i$\\
 $\phi_2=0.3$&0.0625&0.0188$i$&0.152&0.124$i$&0.1050&0.1050$i$&0.253&0.270$i$&0.0700&0.1400$i$\\
 $\phi_2=0.4$&0.0512&0.0107$i$&0.165&0.134$i$&0.1200&0.1200$i$&0.289&0.308$i$&0.0800&0.1600$i$\\
 $\phi_2=0.5$&0.0383&0.0052$i$&0.163&0.130$i$&0.1250&0.1250$i$&0.301&0.321$i$&0.0833&0.1667$i$\\
 $\phi_2=0.6$&&&0.146&0.115$i$&0.1200&0.1200$i$&0.289&0.308$i$&0.0800&0.1600$i$\\
 $\phi_2=0.7$&&&0.116&0.0894$i$&0.1050&0.1050$i$&0.253&0.270$i$&0.0700&0.1400$i$\\
 $\phi_2=0.8$&&&0.0770&0.0575$i$&0.0800&0.0800$i$&0.193&0.206$i$&0.0533&0.1067$i$\\
 $\phi_2=0.9$&&&0.0339&0.0239$i$&0.0450&0.0450$i$&0.108&0.116$i$&0.0300&0.0600$i$\\
 \hline
\end{tabular}
\caption{Coefficients of the $(k_q a)^2$ and $(k_q a)^3$ terms in the
expansion of $A^{(p)}_2$ (given in Eq. (\ref{k-expansion})), for each
of the isotropic models considered here. These are the hard-sphere (HS) model,
fully-penetrable-sphere (FPS) system, Debye random medium (DRM),
random checkerboard (RC), and a power-law-correlated system (PLC).
For hard spheres and fully penetrable spheres, the spheres comprise
phase 2. The parameter $\alpha^{(2)}_j$ is
the $j^{th}$-order coefficient.    As demonstrated in Eq. (\ref{scalarA2}), there are no zeroth
or first order terms in this expansion.}\label{Table:S2}
\end{table}

In Figs. \ref{fig:r1} and \ref{fig:i1}, we show the real and imaginary
components of the effective dielectric constant, respectively, of five
isotropic models described in the previous section, plotted against
dielectric-contrast ratio {at a volume fraction $\phi_2=0.1$.  Since
each model is below or assumed to be below its percolation threshold
at this volume fraction, the two-point approximation
(\ref{two-point-isotropic}) is used in conjunction with
(\ref{k-expansion}) in order to calculate $A_2^{(p)}$.  The real and
imaginary parts of the effective dielectric constant at volume
fraction $\phi_2=0.5$ are plotted in Figs. \ref{fig:r5} and
\ref{fig:i5}, respectively.  At this volume fraction, all structures
except for the hard-sphere model are above their percolation
thresholds.  As shown in Fig. \ref{fig:i1}, the imaginary component of
the effective dielectric constant of the hard-sphere model is
significantly smaller than that of the fully-penetrable-sphere model.
This makes intuitive sense because the constraint that hard spheres
may not overlap results in larger spatial correlations (smaller coarseness), leading
to a smaller imaginary component.  Upon increasing volume fraction,
the hard-sphere model becomes less and less coarse. Close to its
freezing point, at $\phi_2\approx 0.5$, the resulting imaginary
component of $\varepsilon_e$ is extremely small, as shown in
Fig. \ref{fig:i5}.  At volume fraction $\phi_2=0.5$, the real part of
$\varepsilon_e$ for the hard sphere model is significantly smaller
than that of the other models, since it is the only one that is not
percolating.  As Fig. \ref{fig:i5} shows, the imaginary component of
the effective dielectric constant is substantially more sensitive to
microstructure than the corresponding real component of
$\varepsilon_e$.

\begin{figure}[H]
\begin{center}
\includegraphics[width=4.0in]{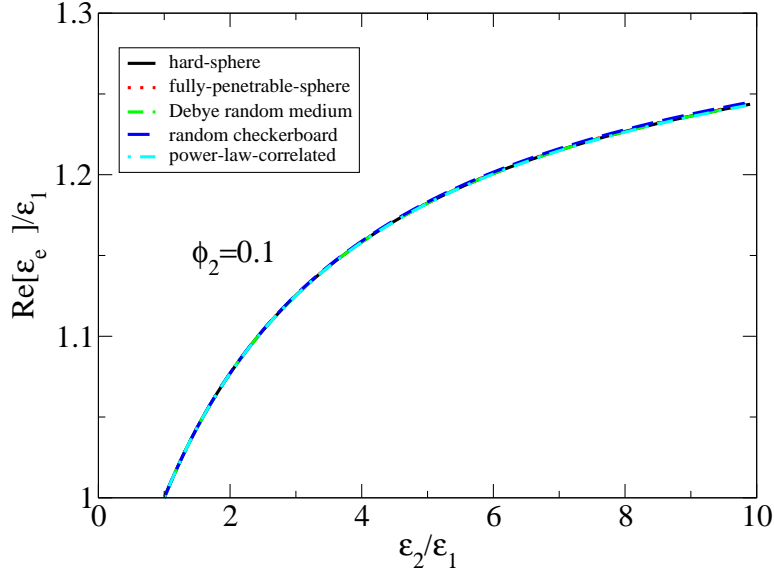}
\caption{(Color online) Real part of the effective dielectric constant
of the various isotropic models studied here as a function
of dielectric-contrast ratio $\varepsilon_2/\varepsilon_1$  at volume fraction $\phi_2=0.1$ and wave
number $k_1=2\pi/(60a)$.   For all models, the two-point
approximation (\ref{two-point-isotropic}) with $p=2$ and $q=1$ is used
since they are each below or assumed to be below their percolation
thresholds.  In order to calculate $A_2^{(p)}$ for each
microstructure, Eq. (\ref{k-expansion}) is used.} \label{fig:r1}
\end{center}
\end{figure}

\begin{figure}[H]
\begin{center}
\includegraphics[width=4in]{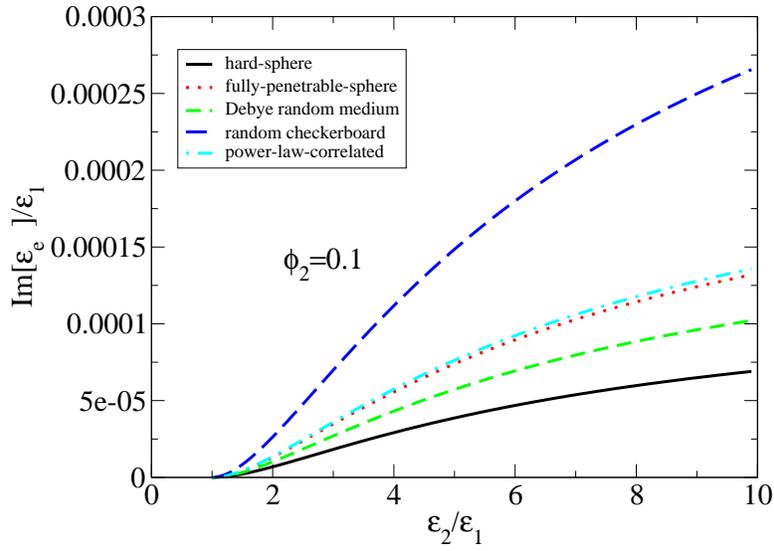}
\caption{(Color online) Corresponding imaginary part of the effective dielectric
constant for the  isotropic models depicted in Fig. \ref{fig:r1}.} \label{fig:i1}
\end{center}
\end{figure}

\begin{figure}[H]
\begin{center}
\includegraphics[width=3.6in]{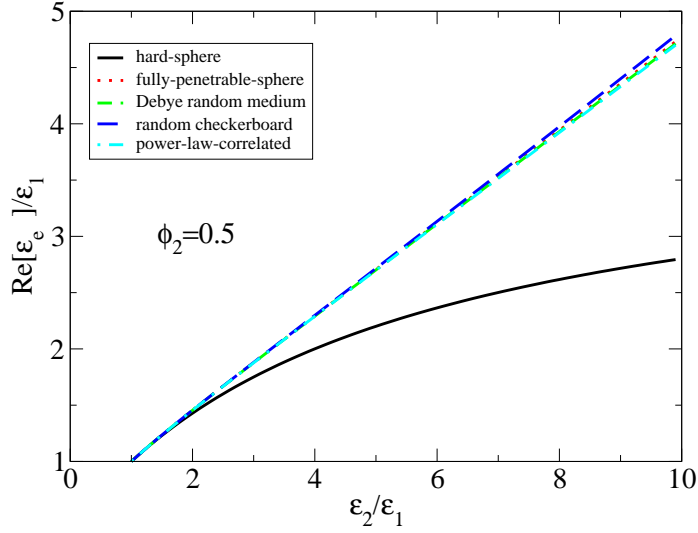}
\caption{(Color online) Real part of the effective dielectric constant
of the various isotropic  models studied here as a
function of dielectric-contrast ratio $\varepsilon_2/\varepsilon_1$ at volume fraction $\phi_2=0.5$
and wave number $k_1=2\pi/(60a)$.   For all of the models besides hard
spheres, the two-point approximation (\ref{two-point-isotropic}) with
$p=1$ and $q=2$ is used, since they are each above their percolation
thresholds.  For the hard-sphere model, which is below
its percolation threshold, we use
(\ref{two-point-isotropic}) with $p=2$ and $q=1$.  Eq. (\ref{k-expansion}) is
used to calculate $A_2^{(p)}$, }\label{fig:r5}
\end{center}
\end{figure}

\begin{figure}[H]
\begin{center}
\includegraphics[width=3.6in]{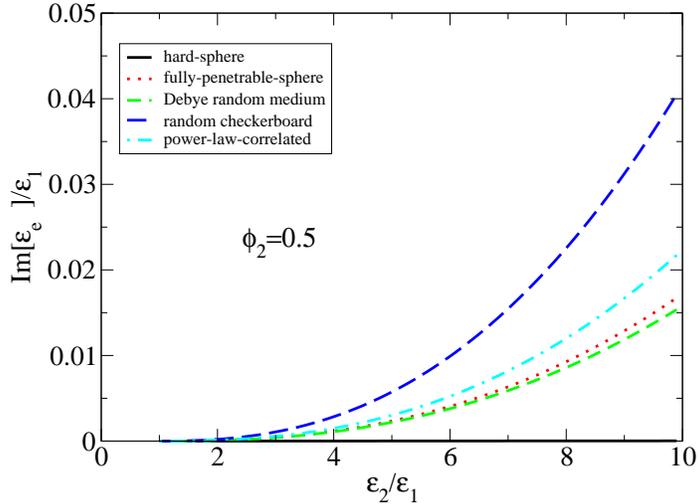}
\caption{(Color online) Corresponding imaginary part of the effective dielectric
constant of the isotropic models depicted in Fig. \ref{fig:r5}.
On the scale of this figures, the hard-sphere curve is almost indistinguishable
from the horizontal axis. }\label{fig:i5}
\end{center}
\end{figure}

In Figs. \ref{fig:rp} and \ref{fig:ip}, we plot the real and imaginary
components of $\varepsilon_e$ for the fully-penetrable-sphere model
as a function of volume fraction.   The two-point approximation
(\ref{two-point-isotropic}) is used, with $p=2$ and $q=1$ for $0\leq
\phi_2 \leq 0.2$ (which is below the percolation threshold of
$\phi_2=0.2895$), and $p=1$ and $q=2$ for $0.4\leq \phi_2 \leq 1$.  A
spline fit is used to interpolate between these two curves to yield an
approximation for $0.2<\phi_2<0.4$.  Eq. (\ref{k-expansion}) is used
to calculate $A_2^{(p)}$.

\begin{figure}[H]
\begin{center}
\includegraphics[width=3.6in]{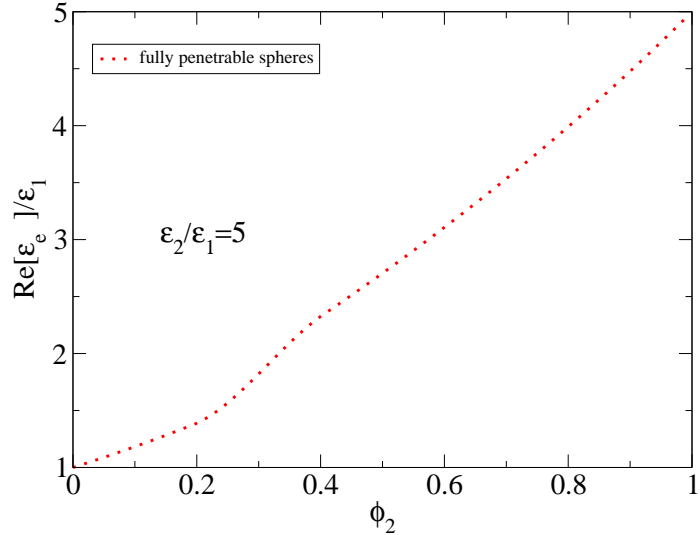}
\caption{(Color online) Real part of the effective dielectric constant
of the fully-penetrable-sphere model as a function of volume
fraction $\phi_2$ of phase 2 at a dielectric contrast ratio $\varepsilon_2/\varepsilon_1=5$ and
at wave number $k_1=2\pi/(60a)$. The two-point approximation
(\ref{two-point-isotropic}) is used with $p=2$ and $q=1$ below the
percolation threshold of $\phi_2=0.2895$, and $p=1$ and $q=2$ above
it.  A spline fit is used to interpolate between these two formulas and
Eq. (\ref{k-expansion}) is used to calculate
$A_2^{(p)}$.}\label{fig:rp}
\end{center}
\end{figure}

\begin{figure}[H]
\begin{center}
\includegraphics[width=3.6in]{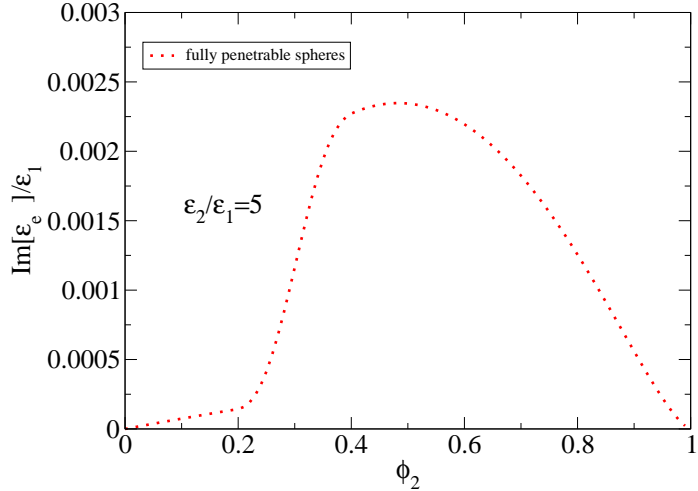}
\caption{(Color online) Corresponding imaginary part of the effective dielectric
constant for the fully-penetrable-sphere model depicted in Fig. \ref{fig:rp}.
A spline fit is used to interpolate between the two formulas. }\label{fig:ip}
\end{center}
\end{figure}

Among the isotropic model microstructures that are discussed here, it
is possible to carry out the full $A^{(p)}_2$ integral as given in
Eq. (\ref{scalarA2}), rather than the expansion given in
Eq. (\ref{k-expansion}).  That said, since for both the real and the
imaginary parts of $A^{(p)}_2$, the next term in the expansion is
smaller by a factor of $(k_q a)^2$, Eq. (\ref{k-expansion}) gives an
extremely good approximation to the effective dielectric constant at
long wavelength.

\subsubsection{Anisotropic Media}

Drawing upon results discussed in Appendix C in which we apply our
formalism at the two-point level to calculate the effective dielectric
tensor of  statistically anisotropic with azimuthal symmetry, we present results for
the axial and planar dielectric constants for an equilibrium dispersion
of hard oriented spheroids in a matrix for a number of spheroid aspect ratios, as predicted by the two-point
approximation, given in Eqs. (\ref{spheroidA2}) and
(\ref{spheroidcalc}).  
Figures \ref{fig:sr} and \ref{fig:si} show real and imaginary
component results, respectively, for the anisotropic hard-spheroid
model, giving axial and in-plane dielectric constants of the
composite in each plot, as predicted by the two-point
approximation, given in Eqs. (\ref{spheroidA2}) and
(\ref{spheroidcalc}).  The former plot, Fig. \ref{fig:sr} shows that
upon increasing the aspect ratio $b/a$,  the axial component of
the dielectric tensor increases.  This makes intuitive sense because in the purely static
case, as the microstructure approaches the long-needle limit, the tensor
component in the axial direction should approach the arithmetic mean
($\sigma_1\phi_1+\sigma_2\phi_2$), which is a rigorous upper bound 
on the effective dielectric constant in the purely static case.
We see that the in-plane component decreases with increasing aspect
ratio.  We may again understand this in the context of the static
regime: in that scenario, the two-point tensor ${\bf A}^{(p)}_2$ must remain
traceless.  The decreasing in-plane component is a direct result of
this property and the fact that the axial component increases
with aspect ratio.  By contrast, we see from Fig. \ref{fig:si} that both the axial
and in-plane imaginary parts of the dielectric tensor increase
with increasing aspect ratio.

\begin{figure}[H]
\begin{center}
\includegraphics[width=4.0in,clip=]{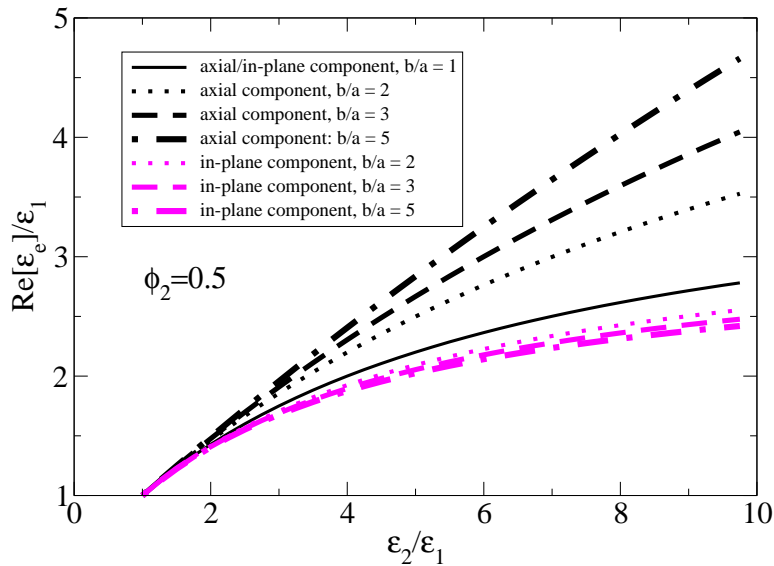}
\caption{(Color online) Real part of the axial and transverse
effective dielectric constants of an equilibrium dispersion of  hard oriented prolate spheroids,
plotted as a function of dielectric-contrast ratio $\varepsilon_2/\varepsilon_1$, for a number of different
aspect ratios $b/a$, as predicted by the two-point approximation, given
in Eqs. (\ref{spheroidA2}) and (\ref{spheroidcalc}).  The contrast is
$\varepsilon_2/\varepsilon_1=5$, the volume fraction is $\phi_2=0.5$, and
$k_1=2\pi/(60a)$, where $a$ is the semi-minor axis of the spheroid.
}\label{fig:sr}
\end{center}
\end{figure}

\begin{figure}[H]
\begin{center}
\includegraphics[width=4.0in]{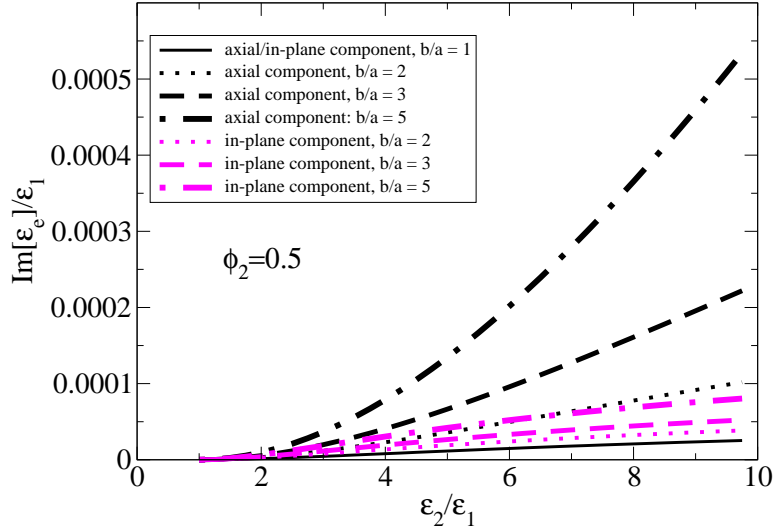}
\caption{(Color online) The corresponding imaginary part of the axial and
transverse effective dielectric constants for the same model
and cases depicted in Fig. \ref{fig:sr}.}\label{fig:si}
\end{center}
\end{figure}

Although we do not show results for the oblate case, we have verified that
as the aspect ratio of the spheroids is increased from the oblate regime ($b/a < 1$) 
through the prolate regime ($b/a > 1$), the real part of the axial dielectric constant increases 
and the real part of the in-plane dielectric constant decreases, which agrees
the behavior given in Ref. \onlinecite{TorquatoLado}.  The imaginary components of both the axial and in-plane dielectric constants
both increase with increasing aspect ratio.

\subsection{Three-Point Estimates}

Here we apply the isotropic three-point
approximation, given explicitly in Eq. (\ref{three-point-isotropic}),
(with $p=1$ and $q=2$) to the fully-penetrable-sphere model in order to ascertain the importance of
three-point information. 
For this model, Table \ref{Table:3pt} gives the coefficients of
$(k_q a)$ for the first two terms of $A_3^{(p)}$, as given in
Eq. (\ref{k-expansion-3pt}), for a number of volume fractions.  We
call the $j^{th}$ order coefficient in this expansion
$\alpha^{(3)}_j$, such that
$A^{(p)}_3=\sum_{j=0}^{\infty}\alpha^{(3)}_j (k_q a)^j$.

\begin{table}
\begin{tabular}{|c||c|c||c|c|}
\hline
 &\multicolumn{2}{c|}{$p=2$,$q=1$}&\multicolumn{2}{c|}{$p=1$,$q=2$}\\
 \cline{2-5}
 &$\alpha^{(3)}_0$&$\alpha^{(3)}_2$&$\alpha^{(3)}_0$&$\alpha^{(3)}_2$\\
 \hline\hline
$\phi_p=        0.1     $       &       0.010   &       0.012   &       0.079   &       0.013   \\
$\phi_p=        0.2     $       &       0.035   &       0.040   &       0.17    &       0.043   \\
$\phi_p=        0.3     $       &       0.070   &       0.075   &       0.24    &       0.085   \\
$\phi_p=        0.4     $       &       0.11    &       0.11    &       0.31    &       0.13    \\
$\phi_p=        0.5     $       &       0.14    &       0.13    &       0.35    &       0.17    \\
$\phi_p=        0.6     $       &       0.17    &       0.15    &       0.37    &       0.20    \\
$\phi_p=        0.7     $       &       0.17    &       0.14    &       0.35    &       0.21    \\
$\phi_p=        0.8     $       &       0.15    &       0.11    &       0.28    &       0.18    \\
$\phi_p=        0.9     $       &       0.10    &       0.053   &       0.17    &       0.12    \\
 \hline
\end{tabular}
\caption{ Coefficients of the $(k_q a)^0$ and $(k_q a)^2$ terms in the
expansion of $A^{(p)}_3$ (given in Eq. (\ref{k-expansion-3pt})), for
the fully-penetrable-sphere model, where the spheres comprise phase
2.  The parameter $\alpha^{(3)}_j$ is the $j^{th}$-order coefficient.
As demonstrated in Eq. (\ref{k-expansion-3pt}), there are no first- or third-
order terms in this expansion.}\label{Table:3pt}
\end{table}

The importance of three-point information at high contrast and volume
fraction is demonstrated in Figs. \ref{fig:k12} and \ref{fig:k13},
which give the $(k_1 a)^2$ and $(k_1 a)^3$ coefficients of the
effective dielectric constant of the fully-penetrable-sphere model
at $\phi_2=0.5$, which is well above its percolation threshold.
The effective dielectric constant is calculated via the two-point and
three-point approximations, given in Eqs. (\ref{two-point-isotropic})
and (\ref{three-point-isotropic}), respectively.  These equations
employ the expansions given in Eqs. (\ref{k-expansion}) and
(\ref{k-expansion-3pt}).  For this model, we have taken $q=2$ and
$p=1$.  Clearly, for both the $(k_1 a)^2$ coefficient (which is real),
and the $(k_1 a)^3$ (which is imaginary), three body information plays
a significant role at high volume fraction and contrast.  We also plot
the imaginary component of the effective dielectric constant for this
model, both with and without the third-order contribution in
Fig. \ref{fig:fpseppimag}.  We see that up to a dielectric-contrast
ratio of roughly 5, the two-point and three-point approximations are
in relatively good agreement, but afterwards they increasingly
diverge.  This serves as a test of the convergence of the series.  We
thus see that up to relatively high contrast ratio (i.e., $\varepsilon_2/\varepsilon_1\approx 5$),
the two-point approximation provides a good estimate, implying that
the remaining terms in the full series expansion (\ref{fullsolution})
are negligibly small.

\begin{figure}[H]
\begin{center}
\includegraphics[width=3.7in]{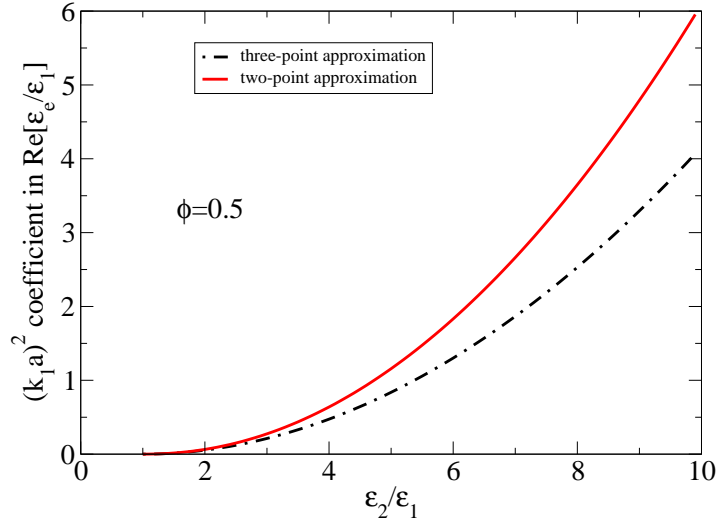}
\caption{(Color online) The coefficient of $(k_1 a)^2$ of the
effective dielectric constant as a function of dielectric-contrast
ratio in the fully-penetrable-sphere model at volume fraction
$\phi_2=0.5$, as predicted by the two-point and three-point
approximations, given in Eqs. (\ref{two-point-isotropic}) and
(\ref{three-point-isotropic}), respectively.   We take $p=1$ and $q=2$
since the system is above its percolation threshold at this volume
fraction.  These equations employ
the expansions given in Eqs. (\ref{k-expansion}) and
(\ref{k-expansion-3pt}).  This term is necessarily real.  The solid
and dashed curves show this quantity with and without the three-point
term included in the expansion given in Eq. (\ref{master}).}\label{fig:k12}
\end{center}
\end{figure}

\begin{figure}[H]
\begin{center}
\includegraphics[width=3.7in]{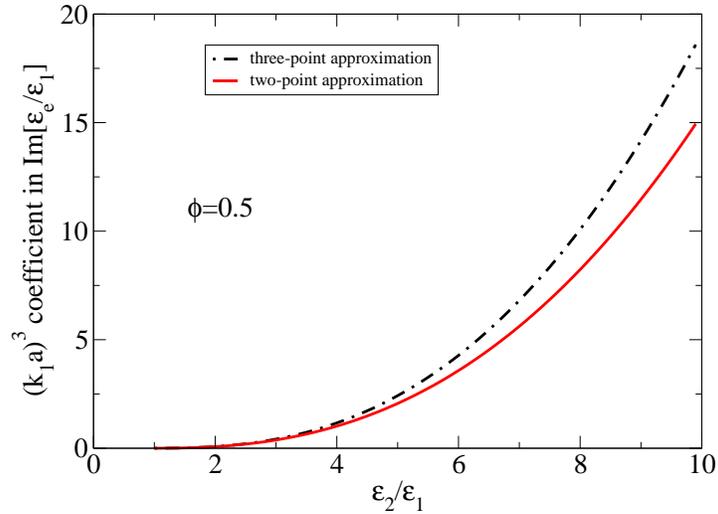}
\caption{(Color online) As in Fig. \ref{fig:k12}, except for the coefficient of $(k_1 a)^3$.}\label{fig:k13}
\end{center}
\end{figure}

\begin{figure}[H]
\begin{center}
\includegraphics[width=4in]{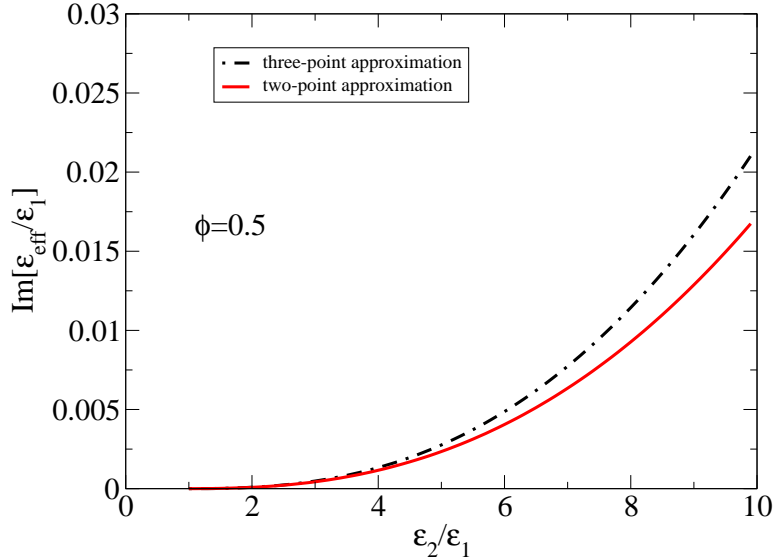}
\caption{(Color online) The imaginary part of the effective
dielectric constant of of the fully-penetrable-sphere model at volume
fraction $\phi_2=0.5$ and wave number $k_1=2\pi/(60a)$, as predicted by
the two-point and three-point approximations, given in
Eqs. (\ref{two-point-isotropic}) and (\ref{three-point-isotropic}),
respectively.   We take $p=1$ and $q=2$ since the system is above
its percolation threshold at this volume fraction.  These equations
employ the expansions given in Eqs. (\ref{k-expansion}) and
(\ref{k-expansion-3pt}).  The solid and dashed curves show this
quantity with and without the three-point term included in the
expansion given in Eq. (\ref{master}).}\label{fig:fpseppimag}
\end{center}
\end{figure}

\section{Conclusions}

We have derived exact strong-contrast expansions (\ref{fullsolution}) for the effective
dielectric tensor $\epeff$ of electromagnetic waves propagating in a
two-phase composite random medium with complex-valued isotropic components 
in the long-wavelength regime. These expansions are not formal but rather  are explicitly given in terms
of certain integrals over the $n$-point correlation functions 
that statistically characterize the medium. To our knowledge,
such an exact representation has not been given explicitly before.
 The nature of the strong-contrast expansion parameter results
 in a radius of convergence of the series (\ref{fullsolution}) that is significantly widened
beyond that of a weak-contrast expansion (i.e., the simple difference of the
dielectric constants of the two phases). Because the expansions
can be considered to be perturbations about the solutions
of the dielectric tensor of certain optimal
structures, we argued that the first few terms of the expansion (\ref{fullsolution}) should
yield a reasonable approximation of  $\epeff$, depending
on whether the high-dielectric phase is below or above
its percolation threshold. In particular, truncations of the exact expansion
led to two- and three-point approximations for $\epeff$, which we applied  to a variety of different
three-dimensional model microstructures, including dispersions of hard
spheres, hard oriented spheroids and  fully penetrable spheres as well as  Debye random media,
random checkerboard, and power-law-correlated materials.

We payed special attention in our analysis to case
in which the components have real dielectric constants but  where the effective dielectric
tensor possesses imaginary components due to disorder, a phenomenon which is essential to applications such as remote
sensing (e.g. in the calculation of the backscattering coefficient \cite{TK}.  In examining two-point approximations of the effective
dielectric constant for  statistically homogeneous and isotropic media
with component phases having purely real dielectric constants, we found
that the imaginary part of   $\epeff$ is related to the
coarseness for very large windows, which may be regarded to be a crude
measure of disorder.  Among other results, we found that the
equilibrium hard-sphere model for  volume fractions up to its freezing
point exhibit a much smaller imaginary component of the effective
dielectric constant than the other four 
statistically homogeneous and isotropic model microstructures  studied here.

For dispersions of fully penetrable spheres, we analyzed the behavior of
the effective dielectric constant using the two- and three-point
approximations, Eqs. (\ref{two-point-isotropic}) and
(\ref{three-point-isotropic}), respectively.  Our results suggest that
truncation of the exact expansion (\ref{fullsolution}) at the
two-point level yields good convergence up to relatively
high phase-contrast ratios ($\approx 5$).  However, the two
approximations increasingly diverge from one another for higher
values, showing the importance of using three-point information at
sufficiently high contrast ratios.

In order to demonstrate the application of our formalism to a
statistically anisotropic and hence macroscopically anisotropic media, we
examined dispersions  of equilibrium hard spheroids.  It was shown that as
the aspect ratio $b/a$ was increased, the real parts of the axial
dielectric constant and the in-plane dielectric constants increased
and decreased, respectively.  The imaginary components of the axial
and in-plane dielectric constants were both found to increase upon
increasing the aspect ratio $b/a$ from 1 (i.e., sphere point).

The dichotomy between periodic media, which have zero imaginary
component in their effective dielectric constants, and disordered
media, which do have such a component, begs an important question:
which is true of quasiperiodic structures?  These structures possess
long-range order but have no translational symmetry, are therefore are
in a sense intermediate between the latter two regimes.  In Appendix
B, we show that at the two-point level, an imaginary component is
obtained only if the Fourier transform of the correlation function is
nonzero at $k_q$.  For two-phase media with periodic structure (i.e.,
crystals), this implies that at this level there is no imaginary
component, but this argument does not hold for media with
quasiperiodic structure (i.e., quasicrystals) \cite{LevineSteinhardt}.
Since the diffraction pattern for quasicrystals possess only
discrete Bragg peaks, the coarseness is necessarily zero.  However,
this does not imply that the imaginary component of  $\epeff$ is
identically zero, since higher-order $k_q$ coefficients may
contribute. The calculation of the effective properties of
quasicrystal two-phase media and a fundamental understanding of their
wave propagation properties remains a challenging open question.

 In future work, we plan on applying the strong-contrast formalism of this
paper  to a number of new model microstructures and explore
other comparison materials in the spirit of Ref. \onlinecite{Ph03} to yield even better
approximations for   $\epeff$.  The procedure presented here is applicable in
any dimension and is thus well suited to the study of scalar Helmholtz
equation for arbitrary space dimension.  One application of this case lies in the
calculation of effective properties of phononic systems.  Another
possible extension involves generalizing analogous elastostatic
results \cite{TorquatoElastic,RHM} to the dynamic case.

\begin{acknowledgments}

The authors are grateful for useful discussions with Paul Chaikin
and Ping Sheng.
This work was supported by the Air Force Office of Scientific Research
under Grant No. F49620-03-1-0406 and the National
Science Foundation under Grant No. DMR-0606415. M. R. 
acknowledges the support of the Natural Sciences and Engineering Research Council of Canada.

\end{acknowledgments}

\appendix

\section{Two-Point Expansion for Arbitrary Comparison Material}

The strong-contrast expansion (\ref{fullsolution}) for the  effective dielectric
tensor was derived for the choice of the comparison material
such that $\varepsilon_0=\varepsilon_q$ ($q=1$ or 2).  This 
this choice significantly simplifies the analysis and has the desirable feature that
it can be regarded to be a solution that perturbs
around the effective dielectric tensor for certain optimal microstructures.
However, for other microstructures, different comparison
materials may offer advantages in terms of better series convergence \cite{RHM,Eyre1}
and more accurate approximations \cite{Ph03}. Here we present 
an equivalent relation to Eq. (\ref{two-point-isotropic}) 
for macroscopically isotropic media but 
for an arbitrary comparison material with dielectric constant $\varepsilon_0$:
\begin{equation}
\frac{1}{\beta_{e0}}=\frac{1}{\beta_{q0}+(\beta_{p0}-\beta_{q0})\phi_p}
-\frac{(\beta_{p0}-\beta_{q0})^2}{\left(\beta_{q0}
+(\beta_{p0}-\beta_{q0})\phi_p\right)^2}A_2^{(p)},
\end{equation}
where $\beta_{e0}=(\varepsilon_e -\varepsilon_0)/(\varepsilon_e +2\varepsilon_0)$,
$\beta_{p0}=(\varepsilon_p -\varepsilon_0)/(\varepsilon_p +2\varepsilon_0)$,
and $\beta_{q0}=(\varepsilon_q -\varepsilon_0)/(\varepsilon_q +2\varepsilon_0)$.
In Ref. \onlinecite{TK}, $\varepsilon_0$ is taken to be the
Bruggeman effective dielectric constant, but this is  not
the best choice for general microstructures \cite{Sheng}. For example,
for the large class of dispersions discussed in Section II.C,
 the choice $\varepsilon_0=\varepsilon_q$ is better.

\section{Proof that $A_2$ is Purely Real at the Two-Point Level}

All of the random media studied here have had dielectric tensors with
non-zero imaginary component. Thus, electromagnetic waves propagating
through these materials will attenuate, albeit with small decay
constants.  Roughly speaking, the physical cause for this attenuation
is the fact that the waves are scattered off of the heterogeneities,
and the scattered waves are no longer coherent with the propagating
wave, and thus, this energy is ``lost'', when homogenization is
applied in Eq. (\ref{homogenization}).  However, this is not so for
periodic heterogeneous media. In the language
of solid-state physics, scattering does not take place because waves
are allowed to propagate coherently as Bloch waves (as opposed to in
the form of pure plane waves).  In this Appendix, we show this
explicitly at the two-point level.    Although a single
periodic configuration is statistically inhomogeneous, 
our formulation (valid for statistically homogeneous media)
can still be applied by first performing a translational
average, i.e.,  averaging over uniformly random displacements
of the origin. This produces averaged quantities,
such as the correlation function defined by Eq. (\ref{correlationfunction}),
translationally invariant. Thus, invoking an ergodic
hypothesis \cite{RHM}, the ensemble average is equal to an
infinite-volume average over the variable $\rv_1$ of a single
(periodic) realization. We limit ourselves here to
periodic media that yield macroscopically isotropic dielectric
tensors.

The two-point coefficient (\ref{scalarA2}) in the macroscopically isotropic case can
be written as follows:
\begin{equation}
A_2^{(p)}=\frac{2k_q^2}{(4\pi)}\int d\rv
\frac{e^{ik_qr}}{r}\left[S^{(p)}_2(\rv)-\phi_p^2\right].
\label{AppendixA2}
\end{equation}
 We define a ``difference indicator function'' as follows:
\begin{equation}
V(\rv) = \Ind^{(p)}(\rv)-\phi_p
\end{equation}
such that $\left(S^{(p)}_2(\rv)-\phi_p^2\right) = \langle
V(\Rv)V(\Rv+\rv)\rangle $, where the average is taken over the dummy
variable $\Rv$.

Representing this real-space correlation function in its spectral
form, we have
\begin{equation}
\langle V(\Rv)V(\Rv+\rv)\rangle = \int \frac{d\qv}{(2\pi)^3} \tilde{V}
(\qv) \tilde{V} (-\qv) \exp (i\qv\cdot\rv).
\end{equation}
where $\tilde{V} (\qv)$ is the Fourier transform of the difference
indicator function at wave vector $\qv$.  Inserting this expression
into Eq. (\ref{AppendixA2}), we have
\begin{eqnarray}
A_2^{(p)}& =& \frac{2k_q^2}{(4\pi)}\int d\rv \frac{e^{ik_qr}}{r}\int
\frac{d\qv}{(2\pi)^3} \tilde V (\qv) \tilde{V} (-\qv) \exp
(i\qv\cdot\rv) \nonumber \\
&=& \frac{2k_q^2}{(4\pi)}\int \frac{d\qv}{(2\pi)^3}
\tilde{V} (\qv) \tilde{V} (-\qv) \int d\rv \exp (i\qv\cdot\rv)
\frac{e^{ik_qr}}{r}.
\end{eqnarray}
We see that the inner integral is just the Fourier transform of the
Green's function of the Helmholtz equation. This inner integral is
thus simply $4\pi/(q^2-k_q^2)$.  Thus, taken together, we have
\begin{eqnarray}
A_2^{(p)} = \frac{2k_q^2}{(4\pi)}\int\frac{d\qv}{(2\pi)^3} \tilde V
(\qv) \tilde{V} (-\qv)\frac{4\pi}{q^2-k_q^2}
\end{eqnarray}

We can see that there are two poles in the last integral located  at $q=\pm k_q$.
Clearly, for a periodic medium, $\tilde V (\qv) \tilde{V} (-\qv)$ is
identically zero at every point, save the reciprocal lattice vectors,
where there are delta functions.  If we assume long wavelength, then
$k_q \ll G$, where $G$ is the magnitude of the smallest non-zero
reciprocal lattice vector.  Thus, the pole will not enter in to the
calculation!  This proves that there can be no imaginary component to
this integral, and this must be the case for all periodic structures.
For the case of a random medium, since there are indeed correlations
at long wavelength (of which there are of course none in the periodic
case), there is a non-zero residue at the pole, and the integral can
be complex.

\section{${\bf A}^{(p)}_2$ for Azimuthal Symmetry}

We now discuss how to explicitly express the tensor 
two-point coefficient  ${\bf A}^{(p)}_2$ [cf. (\ref{two-point-integral})] 
for statistically anisotropic media with an axis of  symmetry (say, the $z$ axis) i.e.,
azimuthal symmetry. An example of such a microstructure is a dispersion of 
hard oriented spheroids, which we discussed in Section III.C.    
Here we follow the methodology of Torquato and Lado \cite{TorquatoLado}, who evaluated similar
integrals for the purely static case.
If we align the Cartesian coordinate system  with the principal
axes frame. the two-point coefficient  ${\bf A}^{(p)}_2$ is diagonal, i.e.,
\begin{equation}
{\bf A}_2^{(p)}=\left(
\begin{array}{lll}
 U & 0 & 0 \\
 0 & U & 0 \\
 0 & 0 & V
\end{array}
\right)\label{spheroidA2}
\end{equation}
where $U$ is the in-plane (i.e., $x$-$y$ plane) component and $V$ is the axial  ($z$ axis) component.
We may explicitly write the two-point tensor as
\begin{equation}
{\bf A}_2^{(p)}=\frac{3}{4\pi}\int
d\rv\exp(ik_qr)\frac{(3-3ik_qr-k_q^2r^2)\rhatrhat +
(-1+ik_qr+k_q^2r^2){\bf
I}}{r^3}\left[S^{(p)}_{2}(\rv)-\phi_p^2\right]
\end{equation}
where we have inserted the dyadic Green's function explicitly.  Retaining
terms through  order $k_q^3$, gives the following explicit expressions for
$U$ and $V$ in spherical coordinates:
\begin{eqnarray}
U &=& \frac{3}{4\pi}\int \frac{d\rv}{r^3} (-1 +
\frac{3}{2}\sin^2(\theta))\left[S^{(p)}_{2}(\rv)-\phi_p^2\right] +
\frac{k_q^2}{2} \frac{3}{4\pi}\int \frac{d\rv}{r}
(1+\frac{1}{2}\sin^2(\theta))\left[S^{(p)}_{2}(\rv)-\phi_p^2\right]\nonumber \\
&&+ \frac{2ik_q^3}{3} \frac{3}{4\pi}\int d\rv
\left[S^{(p)}_{2}(\rv)-\phi_p^2\right]
\label{ueq}
\end{eqnarray}
and
\begin{eqnarray}
V &=& \frac{3}{4\pi}\int \frac{d\rv}{r^3} (-1 +
3\cos^2(\theta))(S^{(p)}_{2}(\rv)-\phi_p^2) + \frac{k_q^2}{2}
\frac{3}{4\pi}\int \frac{d\rv}{r}
(1+\cos^2(\theta))\left[S^{(p)}_{2}(\rv)-\phi_p^2)\right] \nonumber \\
&& + \frac{2ik_q^3}{3} \frac{3}{4\pi}\int d\rv
\left[S^{(p)}_{2}(\rv)-\phi_p^2\right],
\label{veq}
\end{eqnarray}
where we have used the fact that the two in-plane matrix components must
be equal in order to simplify the integrals.  

In Ref. \onlinecite{TorquatoLado}, the first integral in the
expression for $V$ was evaluated analytically for the case of hard oriented spheroids. 
In what follows, we specialize to this model. We evaluate the first
two integrals in each of the relations (\ref{ueq}) and (\ref{veq}) numerically.
Note that the third integral in each expression, which we denote by $U_3$
and $V_3$, respectively, are identical, and may be obtained analytically
as follows:
\begin{eqnarray}
U_3=V_3&=&\frac{2ik_q^3}{3} \frac{3}{4\pi}\int d\rv
\left[S^{(p)}_{2,HS}(\rv;b/a)-\phi_p^2\right]\nonumber \\
& =& \frac{4\pi
ik_q^3}{3}\frac{3}{4\pi}\int_{-1}^{1} d\cos(\theta)\int_0^\infty dr
r^2 \left[S^{(p)}_{2,HS}(\sigma_0(r/\sigma(\theta));1)-\phi_p^2\right] \nonumber \\
&=&ik_q^3\int_{-1}^{1} d\cos(\theta)\int_0^\infty dr r^2
\left[S^{(p)}_{2,HS}(\sigma_0(r/\sigma(\theta));1)-\phi_p^2\right]\nonumber \\
&=&ik_q^3\int_{-1}^{1}
d\cos(\theta)\left[\frac{\sigma(\theta)}{\sigma_0}\right]^3\int_0^\infty
dr r^2 \left[S^{(p)}_{2,HS}(r;1)-\phi_p^2\right] \nonumber \\
&=&
2ik_q^3\frac{b}{a}\int_0^\infty dr r^2
\left[S^{(p)}_{2,HS}(r;1)-\phi_p^2\right],
\label{spheroid-A}
\end{eqnarray}
where we have used the mapping between the two-point function
$S^{(p)}_{2,HS}(\rv;b/a)$ for the hard-spheroid model and $S^{(p)}_{2,HS}(\rv;1)$  for  the corresponding
hard-sphere system described in Section III.C.
Note that the radial integral in the last line of (\ref{spheroid-A}) is 
proportional to the zero-wave-vector of the structure factor of
the hard-sphere system.  For an equilibrium distribution, this may be obtained analytically directly
from the Percus-Yevick approximation, for example. The two-point
approximation to the dielectric tensor is obtained by truncating
the series expansion given in Eq. (\ref{fullsolution}) after the
second term to yield
\begin{equation}
\beta_{pq}^2\phi_p^2({\bf \varepsilon}_{e} -\varepsilon_q{\bf I})^{-1}({\bf
\varepsilon}_{e} +2\varepsilon_q{\bf I})=\phi_p\beta_{pq}{\bf I} - {\bf
A}^{(p)}_2\beta_{pq}^n.\label{spheroidcalc}
\end{equation}


\begin{thebibliography}{22}
\bibitem[{Foo()}]{Footnote} \bibinfo{note}{Note that there exist
hyperuniform systems that are disordered, (e.g., maximally random
jammed sphere packings), and therefore, $C_\infty=0$ does not
necessarily imply the absence of disorder (see Refs. \onlinecite{StillingerTorquato,Donev})}.
\end{thebibliography}
\end{document}